\documentclass[version=preprint]{iacrcc}

\license{CC-by}

\usepackage{algorithm}
\usepackage{algorithmic}
\usepackage{graphicx}
\graphicspath{ {Figures/} }
\usepackage{multirow}

\newcommand{\Fp}{\mathbb{F}_p}

\newcommand{\Zq}{\mathbb{Z}_q}
\newcommand{\Zn}{\mathbb{Z}_n}
\newcommand{\Zsn}{\mathbb{Z}^{\ast}_{n}}

\title[running  = {Fast PC-MM from Unpacked AHE},
       subtitle = {}
      ]{Fast Plaintext-Ciphertext Matrix Multiplication from Additively Homomorphic Encryption}

\addauthor[orcid    = {0009-0008-0508-2902},
           inst     = {1},
           email    = {krishnasai@iisc.ac.in},
           surname  = {Ramapragada}
          ]{Krishna Sai Tarun Ramapragada}

\addauthor[orcid   = {0000-0001-7949-4178},
           inst    = {1},
           email   = {utsav@iisc.ac.in},
           surname = {Banerjee},
           onclick = {https://banerjeeutsav.github.io},
          ]{Utsav Banerjee}

\addaffiliation[ror     = 04dese585,
                street  = {C V Raman Road},
                city    = {Bengaluru},
                state   = {Karnataka},
                postcode= {560012},
                country = {India}
               ]{Electronic Systems Engineering, Indian Institute of Science}

\addfunding[ror=048xjjh50,
            grantid={PMRF}, 
            country={India}]{Ministry of Education}

\genericfootnote{
This work was supported by the Prime Minister's Research Fellowship (PMRF), Ministry of Education, Government of India.
A revised version of this paper was published in the IACR Communications in Cryptology, vol. 2, no. 1 (2025) - DOI: \href{https://dx.doi.org/10.62056/abhey76bm}{10.62056/abhey76bm}
}

\begin{document}

\maketitle


\begin{abstract}
Plaintext-ciphertext matrix multiplication (PC-MM) is an indispensable tool in privacy-preserving computations such as secure machine learning and encrypted signal processing.
While there are many established algorithms for plaintext-plaintext matrix multiplication, efficiently computing plaintext-ciphertext (and ciphertext-ciphertext) matrix multiplication is an active area of research which has received a lot of attention.
Recent literature have explored various techniques for privacy-preserving matrix multiplication using fully homomorphic encryption (FHE) schemes with ciphertext packing and Single Instruction Multiple Data (SIMD) processing.
On the other hand, there hasn't been any attempt to speed up PC-MM using unpacked additively homomorphic encryption (AHE) schemes beyond the schoolbook method and Strassen's algorithm for matrix multiplication.
In this work, we propose an efficient PC-MM from unpacked AHE, which applies Cussen's compression-reconstruction algorithm for plaintext-plaintext matrix multiplication in the encrypted setting.
We experimentally validate our proposed technique using a concrete instantiation with the additively homomorphic elliptic curve ElGamal encryption scheme and its software implementation on a Raspberry Pi 5 edge computing platform.
Our proposed approach achieves up to an order of magnitude speedup compared to state-of-the-art for large matrices with relatively small element bit-widths.
Extensive measurement results demonstrate that our fast PC-MM is an excellent candidate for efficient privacy-preserving computation even in resource-constrained environments.
\end{abstract}

\begin{textabstract}
Plaintext-ciphertext matrix multiplication (PC-MM) is an indispensable tool in privacy-preserving computations such as secure machine learning and encrypted signal processing.
While there are many established algorithms for plaintext-plaintext matrix multiplication, efficiently computing plaintext-ciphertext (and ciphertext-ciphertext) matrix multiplication is an active area of research which has received a lot of attention.
Recent literature have explored various techniques for privacy-preserving matrix multiplication using fully homomorphic encryption (FHE) schemes with ciphertext packing and Single Instruction Multiple Data (SIMD) processing.
On the other hand, there hasn't been any attempt to speed up PC-MM using unpacked additively homomorphic encryption (AHE) schemes beyond the schoolbook method and Strassen's algorithm for matrix multiplication.
In this work, we propose an efficient PC-MM from unpacked AHE, which applies Cussen's compression-reconstruction algorithm for plaintext-plaintext matrix multiplication in the encrypted setting.
We experimentally validate our proposed technique using a concrete instantiation with the additively homomorphic elliptic curve ElGamal encryption scheme and its software implementation on a Raspberry Pi 5 edge computing platform.
Our proposed approach achieves up to an order of magnitude speedup compared to state-of-the-art for large matrices with relatively small element bit-widths.
Extensive measurement results demonstrate that our fast PC-MM is an excellent candidate for efficient privacy-preserving computation even in resource-constrained environments.
\end{textabstract}


\section{Introduction}
\label{sec:introduction}

Matrix multiplication is a cornerstone of linear algebra, indispensable for representing and computing transformations, relationships and interactions in various fields. It lies at the foundations of various important and interesting applications in machine learning, signal processing, cryptography, finance, robotics, bioinformatics and scientific computing. Its widespread significance is driven by its efficiency, enabling complex computations in diverse domains through advancements in algorithms, software and hardware.

The advent of cloud services, edge computing and Internet of Things (IoT) has raised various privacy concerns because sensitive data remains encrypted only during communication and storage but not during computation. This has led to an important emerging field of research -- \textit{privacy-preserving computation} or \textit{secure outsourced computation}.
Homomorphic encryption (HE) is one of the most promising cryptographic primitives which enables privacy-preserving computation on encrypted data without requiring decryption \cite{acar_homomorphic_2018}.
Partially homomorphic encryption (PHE) schemes, such as the Rivest-Shamir-Adleman (RSA) \cite{rsa_pke_1978}, Goldwasser-Micali (GM) \cite{gm_pke_1982}, ElGamal \cite{elgamal_pke_1985} and Paillier \cite{paillier_pke_1999} cryptosystems, support the evaluation of only one type of operation (additions or multiplications) on ciphertexts.
Fully homomorphic encryption (FHE) schemes, such as the Gentry \cite{gentry_fhe_2009}, Brakerski-Gentry-Vaikuntanathan (BGV) \cite{bgv_fhe_2014}, Brakerski-Fan-Vercauteren (BFV) \cite{bfv_fhe_2012}, Gentry-Sahai-Waters (GSW) \cite{gsw_fhe_2013} and Cheon-Kim-Kim-Song (CKKS) \cite{ckks_fhe_2017} cryptosystems, support the evaluation of both addition and multiplication operations on ciphertexts.
The PHE schemes are based on the hardness of number theoretic problems such as integer factorization, quadratic residuosity, composite residuosity and discrete logarithms, while the FHE schemes are based on the hardness of lattice problems such as learning with errors (LWE) and learning with errors over rings (Ring-LWE).

In the context of HE, plaintext-ciphertext matrix multiplication (\texttt{PC-MM}) is a tool which enables the multiplication of an unencrypted matrix with an encrypted matrix, allowing secure computation on sensitive data without decryption.
\texttt{PC-MM} outputs the ciphertext(s) corresponding to the (encrypted) matrix $\boldsymbol{A} \times \boldsymbol{B}$, where its inputs are the plaintext (unencrypted) matrix $\boldsymbol{A}$ and the ciphertext(s) corresponding to the (encrypted) matrix $\boldsymbol{B}$.
\texttt{PC-MM} is a crucial component of privacy-preserving machine learning which requires computing numerous large matrix products in various encrypted neural network stages such as convolution and fully-connected layers. For example, privacy-preserving inference with deep neural networks (DNNs) and transformer networks for large language models (LLMs) involve plaintext weight matrices and ciphertext user data matrices. While there are many established algorithms for plaintext-plaintext matrix multiplication \cite{clrs_algo_2009}, efficiently computing plaintext-ciphertext (and ciphertext-ciphertext) matrix multiplication is an active area of research which has recently received a lot of attention.

\textbf{Prior Work:}
Constructions of secure matrix multiplication from lattice-based FHE schemes heavily exploit ciphertext packing \cite{brakerski_packed_2013} to accommodate multiple plaintexts in a single ciphertext (in the form of slots) and Single Instruction Multiple Data (SIMD) homomorphic operations (which enable massively parallel execution) to efficiently perform encrypted computations.
Various techniques for FHE-based  plaintext-ciphertext (and ciphertext-ciphertext) matrix multiplication have been proposed by \cite{halevi_helib_2014, lu_matrix_2017, jiang_matrix_2018, wang_matrix_2019, jang_matrix_2022, rizomiliotis_matrix_2022, huang_matrix_2023, zhu_matrix_2023, zheng_matrix_2023, huang_recommender_2023, bae_matrix_2024, aikata_matrix_2024, gao_secure_2024, hong_genotype_2024, ma_pca_2024, park_ccmm_2025} using a combination of homomorphic multiplications, additions and rotations with varying degrees of ciphertext packing.
FHE-based \texttt{PC-MM} with packed ciphertexts and SIMD-style arithmetic computations has been used to demonstrate privacy-preserving applications such as deep learning \cite{phong_deep_2017, juvekar_gazelle_2018, chen_malicious_2020}, federated principal component analysis \cite{froelicher_federated_2023}, smart contracts \cite{liu_verifiable_2023} and transformer inference \cite{hao_iron_2022, ding_east_2023, pang_bolt_2024, zhang_transformer_2024, moon_thor_2024}.
Apart from FHE, secure matrix-vector arithmetic and inner product computations have been explored by \cite{abdalla_ipe_2015, bishop_ipe_2015, agrawal_ipe_2016, datta_ipe_2016, ligier_ipe_2017, kim_ipe_2018, ryffel_ipe_2019, marc_ipe_2019, ahsan_ants_2022, chen_quadratic_2023, banerjee_globecom_2023} from discrete logarithm-based and pairing-based cryptosystems which lack the ciphertext packing and SIMD processing capabilities of FHE. It is important to note that these implementations, especially using FHE-based schemes, face several challenges in terms of computation, communication and memory cost which limit their practical deployment in resource-constrained environments such as the IoT.

In this work, we are particularly interested in the efficient realization of \texttt{PC-MM} from non-lattice-based additively homomorphic PHE schemes which do not support packed ciphertexts.
In the absence of packing, \texttt{PC-MM} of two $n \times n$ matrices $\boldsymbol{A}$ (unencrypted) and $\boldsymbol{B}$ (encrypted) requires $n^2$ ciphertexts to represent the elements of $\boldsymbol{B} = [b_{ij}]_{n \times n}$ \cite{jiang_matrix_2018}.
The schoolbook method \cite{clrs_algo_2009} for matrix multiplication using this setup with an additively homomorphic encryption scheme requires $n^3$ plaintext-ciphertext multiplications and $n^2 (n-1)$ ciphertext-ciphertext additions \cite{ahsan_ants_2022, liu_verifiable_2023}, that is, $O(n^3)$ computational complexity.
This can be improved to $O(n^{\text{log}_2 7}) \approx O(n^{2.807})$ using Strassen's algorithm \cite{strassen_matmul_1969} which reduces the required number of plaintext-ciphertext multiplications at the cost of slight increase in the required number of ciphertext-ciphertext additions and significant increase in the memory requirement \cite{liu_verifiable_2023}. Further mathematical details are provided in Section \ref{subsec:design:traditional}. Beyond the schoolbook method and Strassen's algorithm, there hasn't been any attempt in recent literature to speed up \texttt{PC-MM} using unpacked additively homomorphic encryption. Furthermore, software evaluation of \texttt{PC-MM} in recent work has been mostly restricted to high-performance desktop and server-scale processors, and efficient implementations suitable for edge computing IoT platforms are largely unexplored. We endeavour to address these research gaps in this work.

\textbf{Our Contributions:}
We propose an efficient approach to fast \texttt{PC-MM} from unpacked additively homomorphic encryption (\texttt{AHE}).
Our key observation is that plaintext-ciphertext multiplications are significantly more expensive to compute than ciphertext-ciphertext additions in typical unpacked \texttt{AHE} schemes, so trading off plaintext-ciphertext multiplications for ciphertext-ciphertext additions can be beneficial.
Our proposed \texttt{PC-MM} technique is based on an extension of Cussen's compression-reconstruction algorithm for plaintext-plaintext matrix multiplication \cite{cussen_matmul_2023} to the encrypted setting.
We provide a concrete instantiation of our fast \texttt{PC-MM} using the well-known additively homomorphic elliptic curve ElGamal encryption scheme.
We experimentally validate our proposed approach using a software implementation based on the open-source MIRACL cryptographic library \cite{scott_ecciot_2020}. Our implementation is extensively profiled on a Raspberry Pi 5 edge computing IoT platform, and we provide performance analysis for a wide range of matrix dimensions $n \in \{2^3, 2^4, 2^5, 2^6, 2^7, 2^8, 2^9\}$ and matrix element bit-widths $t \in \{4, 8, 12, 16\}$.
Our measurement results indicate up to an order of magnitude speedup with our proposed \texttt{PC-MM} compared to Strassen's algorithm in case of large matrices with relatively small element bit-widths.
Such large matrices with constrained elements are quite common in practical applications like machine learning, thus making our fast \texttt{PC-MM} an excellent candidate for privacy-preserving computation even in resource-constrained environments.
Finally, we provide a brief discussion on applications of our fast \texttt{PC-MM} technique as well as its extension to another popular unpacked \texttt{AHE} scheme, the Paillier cryptosystem.
\section{Preliminaries}
\label{sec:background}

\begin{figure}[!b]
\centering
\includegraphics[width=0.9\textwidth]{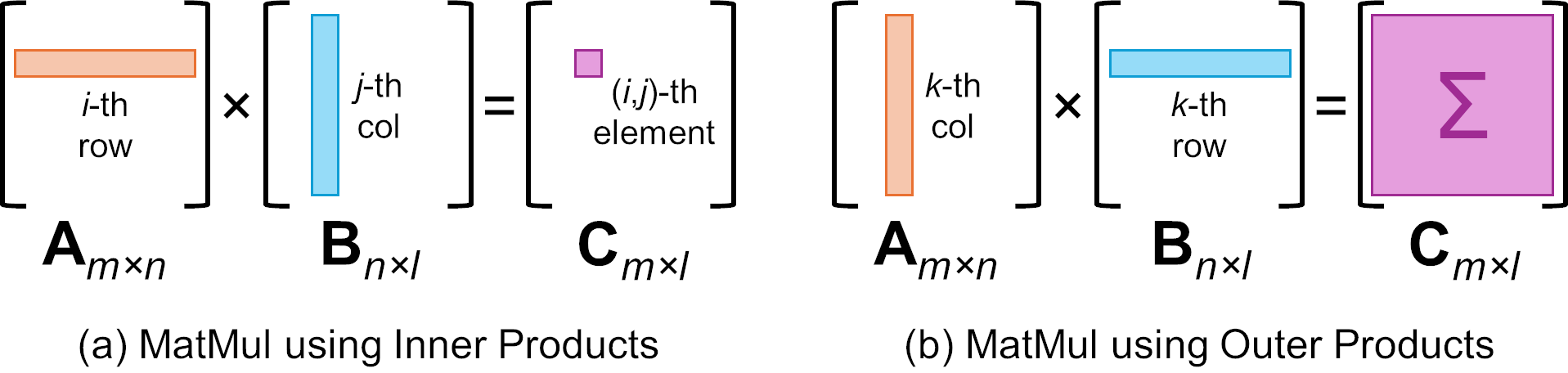}
\caption{Matrix multiplication $\boldsymbol{A}_{m \times n} \times \boldsymbol{B}_{n \times l} = \boldsymbol{C}_{m \times l}$ using (a) row-and-column inner products and (b) column-and-row outer products (diagram inspired by \cite{mlwiki_matmul}).}
\label{fig:matmul}
\end{figure}


We denote matrices and vectors by bold uppercase and bold lowercase letters respectively. For example, an $m \times n$ matrix $\boldsymbol{A}$ is written as $\boldsymbol{A} = [a_{ij}]_{m \times n}$, where $a_{ij}$ is the $(i,j)$-th element with $1 \le i \le m, 1 \le j \le n$. Similarly, an $n \times 1$ column vector $\boldsymbol{b}$ is written as $\boldsymbol{b} = [b_{i}]_{n}$, where $b_{i}$ is the $i$-th element with $1 \le i \le n$. The $i$-th row and $j$-th column of an $m \times n$ matrix $\boldsymbol{A}$ are denoted by $\boldsymbol{A}[i,:]$ and $\boldsymbol{A}[:,j]$ respectively.
Matrix multiplications are denoted by $\times$.
For integer $n$, let $\Zn$ denote the set of integers modulo $n$ and let $\Zsn$ denote the multiplicative group of integers modulo $n$ that are co-prime to $n$.
Let $\Fp$ denote a finite field whose characteristic is a large prime $p$.

\clearpage

\subsection{Matrix Multiplication}
\label{subsec:background:matmul}

We provide a quick refresher on two  matrix multiplication (\texttt{MatMul}) techniques -- using inner products and outer products (as shown in Figure \ref{fig:matmul}):
\begin{itemize}
\item \textbf{\texttt{MatMul} using Row-and-Column Inner Products:} This is the most commonly used matrix multiplication algorithm where each element of the output matrix is computed as an inner product of a row of the first input matrix and a column of the second input matrix. For example, for $m \times n$ and $n \times l$ input matrices $\boldsymbol{A}$ and $\boldsymbol{B}$ respectively, the elements of the $m \times l$ output matrix $\boldsymbol{C} = \boldsymbol{A} \times \boldsymbol{B}$ are calculated as:
\[
c_{ij} = \boldsymbol{A}[i,:] \times \boldsymbol{B}[:,j] = \sum_{k = 1}^{n} a_{ik}b_{kj}
\]
that is, the $(i,j)$-th element $c_{ij}$ of $\boldsymbol{C}$ is the inner product of the $i$-th row of $\boldsymbol{A}$ and the $j$-th column of $\boldsymbol{B}$ ($1 \le i \le m$ and $1 \le j \le l$).
\item \textbf{\texttt{MatMul} using Column-and-Row Outer Products:} This is another well-known matrix multiplication algorithm where the output matrix is computed as a sum of outer products of the columns of the first input matrix and the rows of the second input matrix. For example, for $m \times n$ and $n \times l$ input matrices $\boldsymbol{A}$ and $\boldsymbol{B}$ respectively, the $m \times l$ output matrix $\boldsymbol{C} = \boldsymbol{A} \times \boldsymbol{B}$ is calculated as:
\[
\boldsymbol{C} = \sum_{k = 1}^{n} \boldsymbol{A}[:,k] \times \boldsymbol{B}[k,:]
\]
that is, $\boldsymbol{C}$ is the sum of the outer products of the $k$-th columns of $\boldsymbol{A}$ and the $k$-th rows of $\boldsymbol{B}$ ($1 \le k \le n$).
\end{itemize}

\subsection{Additively Homomorphic Encryption}
\label{subsec:background:ahe}

Consider a \textit{public key encryption} scheme \texttt{PKE} = (\texttt{KeyGen}, \texttt{Encrypt}, \texttt{Decrypt}). Let $\mathcal{M}$ and $\mathcal{C}$ denote the plaintext and ciphertext spaces respectively. Then, standard definitions of the three constituent algorithms of this encryption scheme \texttt{PKE} are as follows \cite{menezes_handbook_2018}:
\begin{itemize}
\item \texttt{KeyGen} ($1^\lambda$): this algorithm generates a public key $pk$ and a secret key $sk$ for the specified security parameter $\lambda$.
\item \texttt{Encrypt} ($pk$, $m$): this algorithm outputs the ciphertext $c \in \mathcal{C}$ corresponding to the input plaintext message $m \in \mathcal{M}$ using the public key $pk$.
\item \texttt{Decrypt} ($sk$, $c$): this algorithm outputs the plaintext message $m \in \mathcal{M}$ corresponding to the input ciphertext $c \in \mathcal{C}$ using the secret key $sk$.
\end{itemize}
The encryption scheme \texttt{PKE} is \textit{correct} if $\texttt{Decrypt} (sk, \texttt{Encrypt} (pk, m)) = m$ for all $m \in \mathcal{M}$ with overwhelming probability.
Now, assume that the message space $\mathcal{M}$ is an additive group with the traditional $+$ operation and the ciphertext space $\mathcal{C}$ also forms a group with an appropriate operation $\oplus$.
Then, this scheme is considered \textit{additively homomorphic} if it satisfies the following property for ciphertexts $c_1$ and $c_2$:
\begin{equation}
c_1 \oplus c_2 = \texttt{Encrypt} (pk, m_1) \oplus \texttt{Encrypt} (pk, m_2) = \texttt{Encrypt} (pk, m_1 + m_2)
\label{eq:ahe_1}
\end{equation}
for any pair of messages $m_1, m_2 \in \mathcal{M}$.
Note that the ciphertext group operation $\oplus$ does not necessarily need to be a numerical addition, e.g., $\oplus$ is a point addition in case of the elliptic curve ElGamal encryption scheme \cite{elgamal_pke_1985} and $\oplus$ is a modular multiplication in case of the Paillier encryption scheme \cite{paillier_pke_1999}, as discussed later.
Henceforth, we will denote such an additively homomorphic public key encryption scheme by \texttt{AHE} = (\texttt{KeyGen}, \texttt{Encrypt}, \texttt{Decrypt}) which is also comprised of similar key generation, encryption and decryption functions as described above. Note that only additively homomorphic encryption schemes which do not support ciphertext packing are of interest in this work.
Then, for such a generic \texttt{AHE} scheme with any ciphertexts $c, c_1, c_2, c_3, \cdots \in \mathcal{C}$, we can define the following:
\begin{equation*}
c_1 \oplus c_2 \oplus c_3 \oplus \cdots = \bigoplus_{i} c_i \,\,\, \text{and} \,\,\, \underbrace{c \oplus c \oplus c \oplus \cdots \oplus c}_{(s-1) \,\, \text{group operations}} = \bigoplus^{s} c
\end{equation*}
Note that the above two ciphertext operations are equivalent to point additions and scalar point multiplication respectively in case of the elliptic curve ElGamal encryption scheme \cite{elgamal_pke_1985}, and modular multiplications and modular exponentiation in case of the Paillier encryption scheme \cite{paillier_pke_1999}, as will be discussed later.

\noindent Clearly, Equation \ref{eq:ahe_1} can be extended to multiple ciphertexts $c_i \in \mathcal{C}$ as:
\begin{equation}
\bigoplus_i c_i = \bigoplus_i \texttt{Encrypt} (pk, m_i) = \texttt{Encrypt} (pk, \sum_i m_i) \,\,\, \forall \,\,\, m_i \in \mathcal{M}
\label{eq:ahe_2}
\end{equation}
Further, Equation \ref{eq:ahe_1} can also be extended to plaintext scalar $s$ and ciphertext $c \in \mathcal{C}$ as:
\begin{equation}
\bigoplus^{s} c = \bigoplus^{s} \texttt{Encrypt} (pk, m) = \texttt{Encrypt} (pk, s \cdot m) \,\,\, \forall \,\,\, m, s \in \mathcal{M} 
\label{eq:ahe_3}
\end{equation}
Equations \ref{eq:ahe_2} and \ref{eq:ahe_3} can be combined to obtain the following generalization of the additively homomorphic property with plaintext scalars $s_i$ and ciphertexts $c_i$:
\begin{equation}
\bigoplus_{i} \bigoplus^{s_i} c_i = \bigoplus_{i} \bigoplus^{s_i} \texttt{Encrypt} (pk, m_i) = \texttt{Encrypt} (pk, \sum_i s_i \cdot m_i) \,\,\, \forall \,\,\, m_i, s_i \in \mathcal{M}
\label{eq:ahe_4}
\end{equation}
Note that this corresponds to homomorphic evaluation of the inner product of vectors $\boldsymbol{s} = (s_1, s_2, \cdots)$ and $\boldsymbol{m} = (m_1, m_2, \cdots)$, where the former is unencrypted and the latter is in the encrypted domain.
This property is traditionally used to realize plaintext-ciphertext matrix multiplication with such an encryption scheme \texttt{AHE}, as discussed in Section \ref{sec:design}.
Finally, Equation \ref{eq:ahe_4} leads to a stricter requirement for correctness of the additively homomorphic encryption scheme that the following should hold with overwhelming probability:
\begin{equation*}
\texttt{Decrypt} (sk, \bigoplus_{i} \bigoplus^{s_i} \texttt{Encrypt} (pk, m_i)) = \sum_i s_i \cdot m_i \,\,\, \text{for} \,\,\, m_i, s_i \in \mathcal{M}
\end{equation*}

\subsection{Elliptic Curve Cryptography}
\label{subsec:background:ecc}

An elliptic curve $E$ over a finite field $\mathbb{K}$ is defined as $E: y^2 + a_1xy + a_3y = x^3 + a_2x^2 + a_4x + a_6$, where $a_1, a_2, a_3, a_4, a_6 \in \mathbb{K}$. In this work, we consider elliptic curves over finite fields with characteristic char($\mathbb{K}$) $\ne 2,3$. In particular, we are interested in fields where the characteristic is a large prime $p > 3$, the corresponding field henceforth denoted as $\Fp$.

The fundamental operations in elliptic curve cryptography (ECC) are \textit{point addition} ($R = P + Q$) and \textit{point doubling} ($R = P + P$), where $P, Q, R \in E(\Fp)$. With these operations, the points on the curve $E(\Fp)$ form an abelian group, with the point at infinity $\infty$ serving as the identity element, that is, $P + \infty = \infty + P = P$ for all $P \in E(\Fp)$. The order of this group (number of points in $E(\Fp)$) is $q$ so that $qP = \infty$ for all $P \in E(\Fp)$.
Repeated addition of a point $P$ with itself is called \textit{elliptic curve scalar multiplication} (ECSM). For any scalar $k$, the scalar multiple $kP$ is defined as \cite{hankerson_ecc_2006}
$kP = P + P + \cdots + P$ (naively requires $k-1$ point additions).
This computation is integral to all ECC protocols and also forms the basis of the underlying \textit{elliptic curve discrete logarithm problem}.
One of the well-known techniques for efficiently computing ECSM is the \textit{Montgomery ladder} shown in Algorithm \ref{algo:ecsm}, which requires exactly $t$ point addition and $t$ point doubling operations to compute $kP$ for $t$-bit scalar $k$ \cite{joye_ecsm_2003}.
Further details on elliptic curve cryptography, ECSM algorithms and ECC protocols are available in \cite{hankerson_ecc_2006, blake_ecc_1999, blake_ecc_2005}.

\begin{algorithm}[!t]
\caption{ECSM using Montgomery ladder \cite{joye_ecsm_2003}}
\label{algo:ecsm}
\begin{algorithmic}[1]
\REQUIRE $k = (k_{t-1}, \cdots, k_1, k_0)_2$ and $P \in E(\Fp)$
\ENSURE $kP$
\STATE $R_0 \leftarrow \infty$, $R_1 \leftarrow P$
\FOR{($i = t-1$; $i \ge 0$; $i = i - 1$)}
\STATE $b \leftarrow k_i$
\STATE $R_{1-b} \leftarrow R_1 + R_0$, $R_b \leftarrow 2R_b$
\ENDFOR
\RETURN $R_0$
\end{algorithmic}
\end{algorithm}

\subsection{Elliptic Curve ElGamal Encryption Scheme}
\label{subsec:background:elgamal}

The ElGamal encryption scheme \cite{elgamal_pke_1985} is one of the oldest and widely studied public key encryption schemes based on the hardness of computing discrete logarithms, and it naturally extends to elliptic curve groups. For elliptic curve ElGamal encryption, the algorithms \texttt{KeyGen}, \texttt{Encrypt} and \texttt{Decrypt} from Section \ref{subsec:background:ahe} are defined as follows:
\begin{itemize}
\item \texttt{KeyGen}: outputs public key $pk = (E(\Fp), q, G, H)$ and secret key $sk = x$ for a given cryptographically suitable elliptic curve group $E(\Fp)$ of order $q$ with generator point $G$, where $x$ is sampled uniformly at random from $[1, q-1]$ and $H = xG$
\item \texttt{Encrypt}: encrypts message $m$ using public key $pk = (E(\Fp), q, G, H)$ and outputs ciphertext $c = (C_1, C_2) = (rG, rH + \phi(m))$ where $r$ is sampled uniformly at random from $[1, q-1]$
\item \texttt{Decrypt}: decrypts ciphertext $c = (C_1, C_2)$ using secret key $sk = x$ and outputs message $m = \phi^{-1}(C_2 - xC_1)$
\end{itemize}
Here, $\phi: \mathcal{M} \rightarrow E(\Fp)$ is an invertible function which maps any message $m \in \mathcal{M}$ to a unique point in the elliptic curve group $E(\Fp)$. A simple and commonly used message-to-point map is to use the ECSM operation such that $\phi(m) = mG$. The inverse $\phi^{-1}$ involves computing a discrete logarithm to retrieve $m$ from $mG$, which is computationally intractable for arbitrary $m \in \Zq$ and necessitates a bound on the message. For $|\mathcal{M}| = B \ll q$, the inverse map can be computed in $O(\sqrt{B})$ space and time complexity using algorithms such as the baby-step giant-step \cite{menezes_handbook_2018}. The value of $B$ is determined by the memory capacity of the underlying implementation platform. Clearly, this scheme satisfies the correctness requirements from Section \ref{subsec:background:ahe} as long as the message space is appropriately bounded. The message (plaintext) space is $\mathcal{M} \subset \Zq$ and the ciphertext space is  $\mathcal{C} = E(\Fp) \times E(\Fp)$.

This scheme is additively homomorphic for $\phi(m) = mG$. Consider the ciphertexts $c_i = (C_1^{(i)}, C_2^{(i)}) = (r_iG, r_iH + m_iG)$ corresponding to messages $m_i \in \mathcal{M}$, where $r_i$ are sampled uniformly at random from $[1, q-1]$. Then, for scalars $s_i \in \mathcal{M}$, we have:
\begin{equation}
\begin{split}
\sum_i s_i \cdot c_i = (\sum_i s_i C_1^{(i)}, \sum_i s_i C_2^{(i)}) &= (\sum_i s_i (r_i G), \sum_i s_i (r_i H + m_i G)) \\
&= (\sum_i s_i r_i G, \sum_i s_i r_i H + \sum_i s_i m_i G)) = (C_1^{\star}, C_2^{\star})
\end{split}
\label{eq:ahe_5_elgamal}
\end{equation}
Clearly, $C_2^{\star} - x C_1^{\star} = \sum_i s_i r_i (xG) + \sum_i s_i m_i G - x (\sum_i s_i r_i G) = (\sum_i s_i m_i) G$ since $H = xG$. Hence, $(C_1^{\star}, C_2^{\star})$ is a valid ciphertext for $\sum_i s_i m_i$ under the same encryption / decryption key pair provided $\sum_i s_i m_i \in \mathcal{M}$ and $\phi^{-1} (C_2^{\star} - x C_1^{\star})$ can be efficiently computed.
Therefore, using the elliptic curve ElGamal scheme as an \texttt{AHE} requires $m_i$ and $s_i$ to be restricted to small subsets of $\mathcal{M}$. For example, if $s_i < B_s$ and $m_i < B_m$ for $1 \le i \le n$, then $\sum_{i=1}^{n} s_i m_i < n B_s B_m < B \ll q$ for $\mathcal{M} = \{0, 1, \cdots, B-1\}$.
These conditions can be easily satisfied in case of plaintext-ciphertext matrix multiplication through the choice of appropriate parameters $B$, $B_s < B$ and $B_m < B$. The ElGamal encryption scheme is not known to support ciphertext packing, thus making it an excellent candidate for demonstrating the techniques proposed in this work.

\subsection{Paillier Encryption Scheme}
\label{subsec:background:paillier}

The Paillier encryption scheme \cite{paillier_pke_1999} is another popular public key encryption scheme based on the hardness of computing composite residues. For Paillier encryption, the algorithms \texttt{KeyGen}, \texttt{Encrypt} and \texttt{Decrypt} from Section \ref{subsec:background:ahe} are defined as follows:
\begin{itemize}
\item \texttt{KeyGen}: outputs public key $pk = (n, g)$ and secret key $sk = (\lambda, \mu)$ where $n = pq$ for large primes $p$ and $q$ such that $\text{gcd} \, (pq, (p-1)(q-1)) = 1$, $\lambda = \text{lcm} \, (p-1, q-1)$, g is sampled uniformly at random from $[1, n^2 - 1]$ such that $\text{gcd} \, (n, L(g^\lambda \, \text{mod} \, n^2)) = 1$ for the quotient function $L(x) = \frac{x-1}{n}$ and $\mu = (L(g^\lambda \, \text{mod} \, n^2))^{-1} \, \text{mod} \, n$
\item \texttt{Encrypt}: encrypts message $m$ using public key $pk = (n, g)$ and outputs ciphertext $c = g^m r^n \, \text{mod} \, n^2$ where $r$ is sampled uniformly at random from $[1, n-1]$
\item \texttt{Decrypt}: decrypts ciphertext $c$ using secret key $sk = (\lambda, \mu)$ and outputs message $m = L(c^\lambda \, \text{mod} \, n^2) \, \mu \, \text{mod} \, n$
\end{itemize}
This scheme also satisfies the correctness requirements from Section \ref{subsec:background:ahe}. The message (plaintext) space is $\mathcal{M} = \Zn$ and the ciphertext space is  $\mathcal{C} = \mathbb{Z}^{\ast}_{n^2}$.
This scheme is also additively homomorphic. Consider the ciphertexts $c_i = g^{m_i} {r_i}^n \, \text{mod} \, n^2$ corresponding to messages $m_i \in \mathcal{M}$, where $r_i$ are sampled uniformly at random from $[1, n-1]$. Then, for scalars $s_i \in \mathcal{M}$, we have:
\begin{equation}
\begin{split}
\prod_i {c_i}^{s_i} = \prod_i (g^{m_i} {r_i}^n \, \text{mod} \, n^2)^{s_i} &= \prod_i g^{s_i m_i} {r_i}^{s_i n} \, \text{mod} \, n^2 \\
&= g^{\sum_i s_i m_i} {(\prod_i {r_i}^{s_i})}^n \, \text{mod} \, n^2 = c^{\ast}
\end{split}
\label{eq:ahe_6_paillier}
\end{equation}
Clearly, $L((c^{\ast})^{\lambda} \, \text{mod} \, n^2) \, \mu \, \text{mod} \, n = \sum_i s_i m_i$. Hence, $c^{\ast}$ is a valid ciphertext for $\sum_i s_i m_i$ under the same encryption / decryption key pair provided $\sum_i s_i m_i \in \mathcal{M}$. Therefore, the Paillier scheme can also be used as an \texttt{AHE} and it does not support ciphertext packing.
\section{Plaintext-Ciphertext Matrix Multiplication}
\label{sec:design}

Next, we discuss how the generalized additively homomorphic property of an unpacked \texttt{AHE} scheme (Equation \ref{eq:ahe_4}) can be used to compute the product of a plaintext matrix and a ciphertext matrix.
While both the elliptic curve ElGamal encryption scheme and the Paillier encryption scheme are popular choices for \texttt{AHE}, the former has much smaller ciphertext sizes.
For example, at the 128-bit security level, elliptic curve ElGamal ciphertexts are 128 bytes long ($= 4 \times 32$ bytes for $\lceil \, \log_2 p \,\rceil = 256$) while Paillier ciphertexts are 768 bytes long ($= 2 \times 384$ bytes for $\lceil \, \log_2 n \,\rceil = 3072$) \cite{menezes_handbook_2018}.
Therefore, we choose the the elliptic curve ElGamal (EC-ElGamal) encryption scheme for our implementation because of its smaller key / ciphertext sizes, computational efficiency and availability of fast software libraries optimized for embedded systems. Henceforth, we use the EC-ElGamal scheme as our \texttt{AHE} with both message and ciphertext spaces being additive groups.

\subsection{Traditional Approach to PC-MM from Unpacked AHE}
\label{subsec:design:traditional}

Consider an $m \times n$ plaintext matrix $\boldsymbol{A} = [a_{ik}]_{m \times n}$ and $nl$ ciphertexts $\texttt{Encrypt} (pk, b_{kj})$ corresponding to an $n \times l$ plaintext matrix $\boldsymbol{B} = [b_{kj}]_{n \times l}$ encrypted under an unpacked \texttt{AHE} scheme as defined in Section \ref{subsec:background:ahe}. Then, the schoolbook method for \texttt{PC-MM} computes the ciphertext corresponding to the $(i,j)$-th element of $\boldsymbol{A} \times \boldsymbol{B} = \boldsymbol{C} = [c_{ij}]_{m \times l}$ as:
\[
\texttt{Encrypt} (pk, c_{ij}) = \texttt{Encrypt} (pk, \sum_{k = 1}^{n} a_{ik}b_{kj}) = \sum_{k = 1}^{n} a_{ik} \cdot \texttt{Encrypt} (pk, b_{kj})
\]
where the row-and-column inner product technique from Section \ref{subsec:background:matmul} has been employed.
Clearly, each inner product requires $n$ plaintext-ciphertext multiplications and $n-1$ ciphertext-ciphertext additions, and $ml$ such inner product computations are required.
For the elliptic curve ElGamal instantiation (EC-ElGamal) of the \texttt{AHE} scheme, 1 plaintext-ciphertext multiplication involves 2 ECSMs (each requiring $t$ point doubling and $t$ point addition operations according to Algorithm \ref{algo:ecsm}), and 1 ciphertext-ciphertext addition involves 2 point addition operations, where $t$ denotes the bit-width of the plaintext matrix elements.
The same analysis also holds for the column-and-row outer product technique from Section \ref{subsec:background:matmul}, where the output is computed as:
\begin{align*}
\begin{bmatrix}
\texttt{Encrypt} (pk, c_{11}) & \texttt{Encrypt} (pk, c_{12}) & \cdots & \texttt{Encrypt} (pk, c_{1l}) \\
\texttt{Encrypt} (pk, c_{21}) & \texttt{Encrypt} (pk, c_{22}) & \cdots & \texttt{Encrypt} (pk, c_{2l}) \\
\vdots & \vdots & \ddots & \vdots \\
\texttt{Encrypt} (pk, c_{m1}) & \texttt{Encrypt} (pk, c_{m2}) & \cdots & \texttt{Encrypt} (pk, c_{ml}) \\
\end{bmatrix} \nonumber \\
= \sum_{k = 1}^{n}
\begin{bmatrix}
a_{1k} \cdot \texttt{Encrypt} (pk, b_{k1}) & a_{1k} \cdot \texttt{Encrypt} (pk, b_{k2}) & \cdots & a_{1k} \cdot \texttt{Encrypt} (pk, b_{kl}) \\
a_{2k} \cdot \texttt{Encrypt} (pk, b_{k1}) & a_{2k} \cdot \texttt{Encrypt} (pk, b_{k2}) & \cdots & a_{2k} \cdot \texttt{Encrypt} (pk, b_{kl}) \\
\vdots & \vdots & \ddots & \vdots \\
a_{mk} \cdot \texttt{Encrypt} (pk, b_{k1}) & a_{mk} \cdot \texttt{Encrypt} (pk, b_{k2}) & \cdots & a_{mk} \cdot \texttt{Encrypt} (pk, b_{kl}) \\
\end{bmatrix}
\end{align*}
which also requires total $mln$ plaintext-ciphertext multiplications and $ml(n-1)$ ciphertext-ciphertext additions.
For square matrices with $m = l = n$, this translates to the $O(n^3)$ complexity for \texttt{PC-MM} as mentioned in Section \ref{sec:introduction}.

The key idea of Strassen's algorithm \cite{strassen_matmul_1969} is to partition the matrices $\boldsymbol{A}$, $\boldsymbol{B}$ and $\boldsymbol{C}$ into equally sized block matrices as follows:
\begin{align*}
\boldsymbol{A} =
\begin{bmatrix}
\boldsymbol{A_{11}} & \boldsymbol{A_{12}} \\
\boldsymbol{A_{21}} & \boldsymbol{A_{22}} \\
\end{bmatrix} \nonumber , \,\,\,
\boldsymbol{B} =
\begin{bmatrix}
\boldsymbol{B_{11}} & \boldsymbol{B_{12}} \\
\boldsymbol{B_{21}} & \boldsymbol{B_{22}} \\
\end{bmatrix} \nonumber \,\,\, \text{and} \,\,\,
\boldsymbol{C} =
\begin{bmatrix}
\boldsymbol{C_{11}} & \boldsymbol{C_{12}} \\
\boldsymbol{C_{21}} & \boldsymbol{C_{22}} \\
\end{bmatrix}
\end{align*}
where the block matrices $\boldsymbol{A_{ij}}$, $\boldsymbol{B_{ij}}$ and $\boldsymbol{C_{ij}}$ are of dimensions $\frac{m}{2} \times \frac{n}{2}$, $\frac{n}{2} \times \frac{l}{2}$ and $\frac{m}{2} \times \frac{l}{2}$ respectively. Then, the output matrix is computed as:
\begin{align*}
\begin{bmatrix}
\boldsymbol{C_{11}} & \boldsymbol{C_{12}} \\
\boldsymbol{C_{21}} & \boldsymbol{C_{22}} \\
\end{bmatrix} \nonumber
=
\begin{bmatrix}
\boldsymbol{M_1} + \boldsymbol{M_4} - \boldsymbol{M_5} + \boldsymbol{M_7} & \boldsymbol{M_3} + \boldsymbol{M_5} \\
\boldsymbol{M_2} + \boldsymbol{M_4} & \boldsymbol{M_1} - \boldsymbol{M_2} + \boldsymbol{M_3} + \boldsymbol{M_6} \\
\end{bmatrix}
\end{align*}
where $\boldsymbol{M_1} = (\boldsymbol{A_{11}} + \boldsymbol{A_{22}}) \times (\boldsymbol{B_{11}} + \boldsymbol{B_{22}})$, $\boldsymbol{M_2} = (\boldsymbol{A_{21}} + \boldsymbol{A_{22}}) \times \boldsymbol{B_{11}}$, $\boldsymbol{M_3} = \boldsymbol{A_{11}} \times (\boldsymbol{B_{12}} - \boldsymbol{B_{22}})$, $\boldsymbol{M_4} = \boldsymbol{A_{22}} \times (\boldsymbol{B_{21}} - \boldsymbol{B_{11}})$, $\boldsymbol{M_5} = (\boldsymbol{A_{11}} + \boldsymbol{A_{12}}) \times \boldsymbol{B_{22}}$, $\boldsymbol{M_6} = (\boldsymbol{A_{21}} - \boldsymbol{A_{11}}) \times (\boldsymbol{B_{11}} + \boldsymbol{B_{12}})$ and $\boldsymbol{M_7} = (\boldsymbol{A_{12}} - \boldsymbol{A_{22}}) \times (\boldsymbol{B_{21}} + \boldsymbol{B_{22}})$. This reduces the number of block matrix multiplications to 7 instead of 8 in the schoolbook approach:
\begin{align*}
\begin{bmatrix}
\boldsymbol{C_{11}} & \boldsymbol{C_{12}} \\
\boldsymbol{C_{21}} & \boldsymbol{C_{22}} \\
\end{bmatrix} \nonumber
=
\begin{bmatrix}
\boldsymbol{A_{11}} \times \boldsymbol{B_{11}} + \boldsymbol{A_{12}} \times \boldsymbol{B_{21}} & \boldsymbol{A_{11}} \times \boldsymbol{B_{12}} + \boldsymbol{A_{12}} \times \boldsymbol{B_{22}} \\
\boldsymbol{A_{21}} \times \boldsymbol{B_{11}} + \boldsymbol{A_{22}} \times \boldsymbol{B_{21}} & \boldsymbol{A_{21}} \times \boldsymbol{B_{12}} + \boldsymbol{A_{22}} \times \boldsymbol{B_{22}} \\
\end{bmatrix}
\end{align*}
As a trade-off, the number of block matrix additions / subtractions is increased to 18 instead of 4.
This process of partitioning into block matrices is recursively applied till the sub-matrices are small enough, thus reducing the overall computational complexity (assuming block matrix multiplications are much more expensive than block matrix additions / subtractions).
Strassen's algorithm can be applied easily in the \texttt{PC-MM} setting where the block matrices $\boldsymbol{B_{ij}}$ and $\boldsymbol{C_{ij}}$ are encrypted as ciphertexts, while the block matrices $\boldsymbol{A_{ij}}$ are in plaintext.
Each recursive iteration of \texttt{PC-MM} with Strassen's algorithm involves 7 plaintext-ciphertext block matrix multiplications and 13 ciphertext-ciphertext block matrix additions / subtractions. There are 5 plaintext-plaintext block matrix additions / subtractions required, but these are much cheaper than encrypted computations
and hence have negligible impact on the overall computational cost.
For square matrices with $m = l = n$, this translates to the $O(n^{\text{log}_2 7}) \approx O(n^{2.807})$ asymptotic complexity for \texttt{PC-MM} as mentioned in Section \ref{sec:introduction}.

\subsection{Cussen's Algorithm for Matrix Multiplication}
\label{subsec:design:cussen}

Cussen's compression-reconstruction algorithm for plaintext-plaintext matrix multiplication was proposed by \cite{cussen_matmul_2023} in the context of improving the energy-efficiency of machine learning hardware accelerators. The key idea of Cussen's algorithm is that a matrix multiplication can be reduced to a ``surprisingly small number'' of scalar multiplications at the cost of extra scalar additions by cleverly pre-processing one of the input matrices and then computing the matrix product as a sum of column-and-row outer products.

Here, we provide a brief description of Cussen's algorithm \cite{cussen_matmul_2023}. The outer product approach for multiplying two matrices $\boldsymbol{A} = [a_{ik}]_{m \times n}$ and $\boldsymbol{B} = [a_{kj}]_{n \times l}$, as explained in Section \ref{subsec:background:matmul}, requires computing the sum of $n$ outer products of the form $\boldsymbol{A}[:,k] \times \boldsymbol{B}[k,:]$, where the $k$-th outer product ($1 \le k \le n$) can be written as:
\begin{align}
\boldsymbol{A}[:,k] \times \boldsymbol{B}[k,:] =
\begin{bmatrix}
a_{1k} \cdot b_{k1} & a_{1k} \cdot b_{k2} & \cdots & a_{1k} \cdot b_{kl} \\
a_{2k} \cdot b_{k1} & a_{2k} \cdot b_{k2} & \cdots & a_{2k} \cdot b_{kl} \\
\vdots & \vdots & \ddots & \vdots \\
a_{mk} \cdot b_{k1} & a_{mk} \cdot b_{k2} & \cdots & a_{mk} \cdot b_{kl} \\
\end{bmatrix}
\label{eq:matmul_outer_product_1}
\end{align}
This outer product computation can be further decomposed into vector-scalar multiplications between $[a_{1k}, a_{2k}, \cdots, a_{mk}]^T$ and $b_{kj}$ which result in the $j$-th column of the $k$-th outer product ($1 \le k \le n$ and $1 \le j \le l$) as:
\begin{align}
\begin{bmatrix}
a_{1k} \\
a_{2k} \\
\vdots \\
a_{mk} \\
\end{bmatrix}
\cdot b_{kj} =
\begin{bmatrix}
a_{1k} \cdot b_{kj} \\
a_{2k} \cdot b_{kj} \\
\vdots \\
a_{mk} \cdot b_{kj} \\
\end{bmatrix}
\label{eq:matmul_outer_product_2}
\end{align}
The core of Cussen's algorithm involves compressing $[a_{1k}, a_{2k}, \cdots, a_{mk}]^T$ into a shorter vector through repeated sorting, eliminating duplicates and taking differences between consecutive elements. Then, the vector-scalar multiplication is computed by first multiplying this short constrained vector with $b_{kj}$ and accumulating the differences to reconstruct the final result. Algorithm \ref{algo:cussen} provides an outline of Cussen's iterative compression-reconstruction method for plaintext vector-scalar multiplication \cite{cussen_matmul_2023}.
This method is then extended across the entire matrices $\boldsymbol{A}$ and $\boldsymbol{B}$ to correctly compute their matrix product $\boldsymbol{A} \times \boldsymbol{B}$. Henceforth, we refer to these two stages of Cussen's algorithm as \textit{Compression Phase} and \textit{Reconstruction Phase} respectively.

\begin{algorithm}[!t]
\caption{Cussen's algorithm for plaintext vector-scalar multiplication \cite{cussen_matmul_2023}}
\label{algo:cussen}
\begin{algorithmic}[1]
\REQUIRE plaintext vector $\boldsymbol{a} = [a_1, a_2, \cdots, a_n]^T$ of length $n$, plaintext scalar $C$ and number of iterations $N$ of vector compression and reconstruction
\ENSURE plaintext vector $\boldsymbol{b} = C \cdot \boldsymbol{a} = [C \cdot a_1, C \cdot a_2, \cdots, C \cdot a_n]^T$
\STATE $n_0 \leftarrow n$
\STATE $\boldsymbol{a}^{(0)} \leftarrow \boldsymbol{a}$
\STATE Compression Phase:
\FOR{($w = 1$; $w \le N$; $w = w + 1$)}
\STATE Sort vector $\boldsymbol{a}^{(w - 1)}$ and eliminate duplicates to obtain vector $\boldsymbol{a}^{(w)} = [a^{(w)}_1, a^{(w)}_2, \cdots]^T$ of length $n_w \le n_{w-1}$ and store pointers to track sorting order
\IF{$ w \ne N$}
\STATE Compute differences of consecutive elements of $\boldsymbol{a}^{(w)}$ as:
\FOR{($i = n_w$; $i > 1$; $i = i - 1$)}
\STATE $a^{(w)}_i \leftarrow a^{(w)}_i - a^{(w)}_{i-1}$
\ENDFOR
\ENDIF
\ENDFOR
\STATE Compute $\boldsymbol{b}^{(N)} = [b^{(N)}_1, b^{(N)}_2, \cdots]^T = C \cdot \boldsymbol{a}^{(N)} = [C \cdot a^{(N)}_1, C \cdot a^{(N)}_2, \cdots]^T$ of length $n_N$
\STATE Reconstruction Phase:
\FOR{($w = N$; $w \ge 1$; $w = w - 1$)}
\IF{$ w \ne N$}
\STATE Compute additions of consecutive elements of $\boldsymbol{b}^{(w)}$ as:
\FOR{($i = 1$; $i < n_w$; $i = i + 1$)}
\STATE $b^{(w)}_{i+1} \leftarrow b^{(w)}_{i+1} + b^{(w)}_i$
\ENDFOR
\ENDIF
\STATE Un-sort vector $\boldsymbol{b}^{(w)}$ and re-insert duplicates using tracking pointers to obtain vector $\boldsymbol{b}^{(w-1)} = [b^{(w-1)}_1, b^{(w-1)}_2, \cdots]^T$ of length $n_{w-1} \ge n_{w}$
\ENDFOR
\STATE $\boldsymbol{b} \leftarrow \boldsymbol{b}^{(0)}$
\RETURN $\boldsymbol{b}$
\end{algorithmic}
\end{algorithm}

\begin{figure}[!b]
\centering
\includegraphics[width=0.98\textwidth]{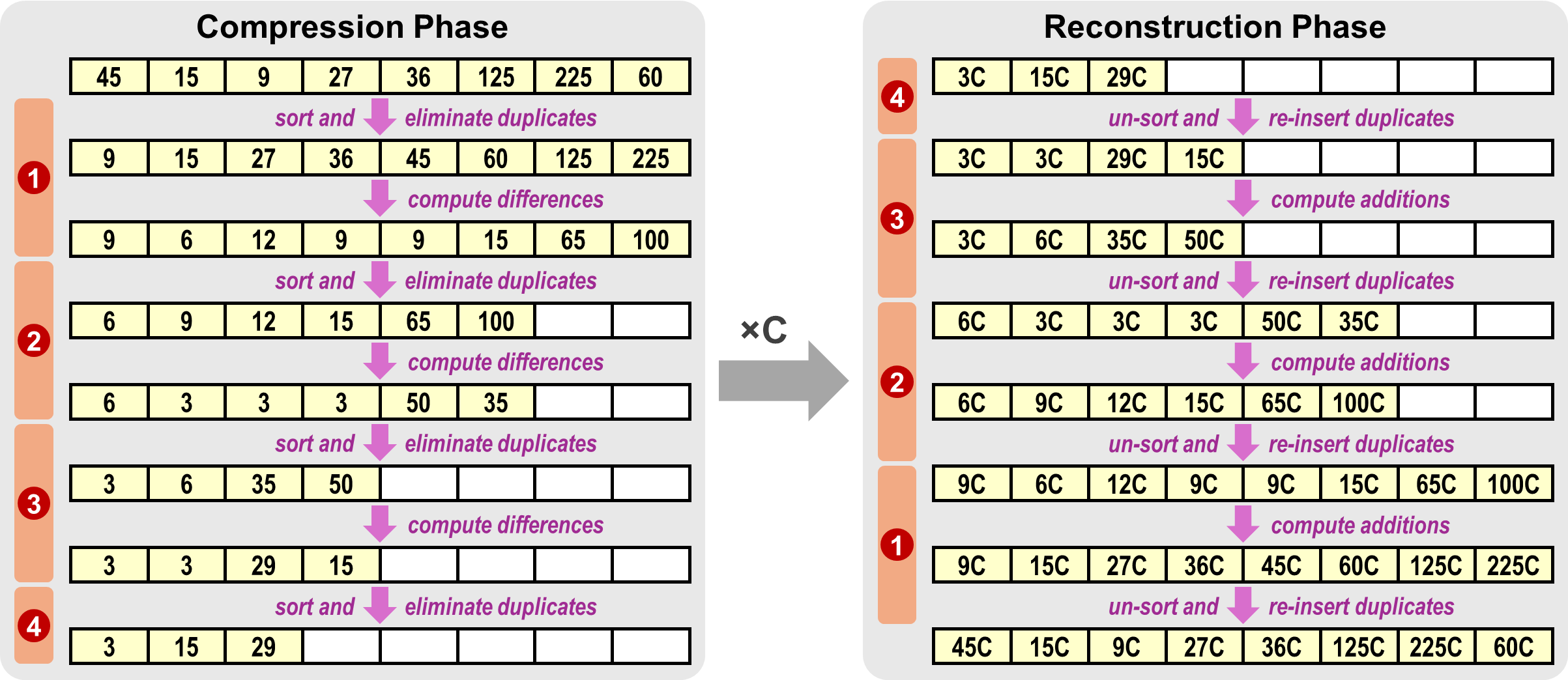}
\caption{Toy example showing four iterations each of the Compression Phase and the Reconstruction Phase of plaintext vector-scalar multiplication using Cussen's algorithm.}
\label{fig:cussen_algo}
\end{figure}

Figure \ref{fig:cussen_algo} shows a toy example of applying $N = 4$ iterations each of the compression and reconstruction phases of Cussen's plaintext vector-scalar multiplication algorithm. This example multiplies a vector of length $n = 8$ with 8-bit elements and a scalar $C$.
The vector is compressed to length 3 and element-wise multiplied by $C$ at the end of the compression phase, where the differences of consecutive vector elements need not be taken in the last iteration. The reconstruction phase then computes the final vector-scalar product by adding back the differences, un-sorting the elements and re-inserting any duplicates using a set of tracking pointers.
Note that this requires 3 multiplications after the compression phase and 15 additions in the reconstruction phase as opposed to 8 multiplications in the schoolbook method, that is, 5 less multiplications and 15 more additions. Therefore, this approach provides an advantage over the schoolbook method if 1 multiplication is more expensive than 3 additions in this toy example of vector-scalar multiplication, ignoring the cost of sorting and tracking duplicates.
In general, for $N$ iterations of Cussen's compression-reconstruction-based plaintext vector-scalar multiplication algorithm, the tracking pointers require $O(N \cdot n)$ storage for vector length $n$ in the worst case.
Further details of the algorithm, its computational complexity, theoretical analysis and various optimizations are available in \cite{cussen_matmul_2023}.

\begin{figure}[!b]
\centering
\includegraphics[width=\textwidth]{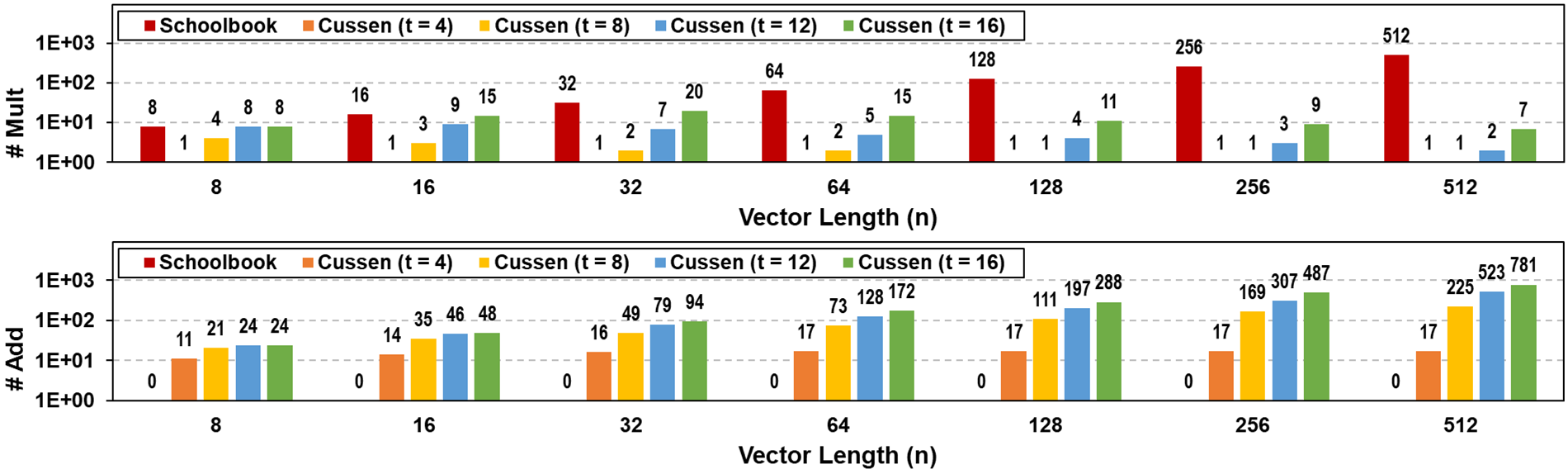}
\caption{Number of multiplications and additions required for plaintext vector-scalar multiplication using schoolbook approach and Cussen's algorithm for random vectors of length $n \in \{2^3, 2^4, \cdots, 2^9\}$ with element bit-widths $t \in \{4, 8, 12, 16\}$.}
\label{fig:cussen_plaintext_vector_python}
\end{figure}

\begin{figure}[!b]
\centering
\includegraphics[width=\textwidth]{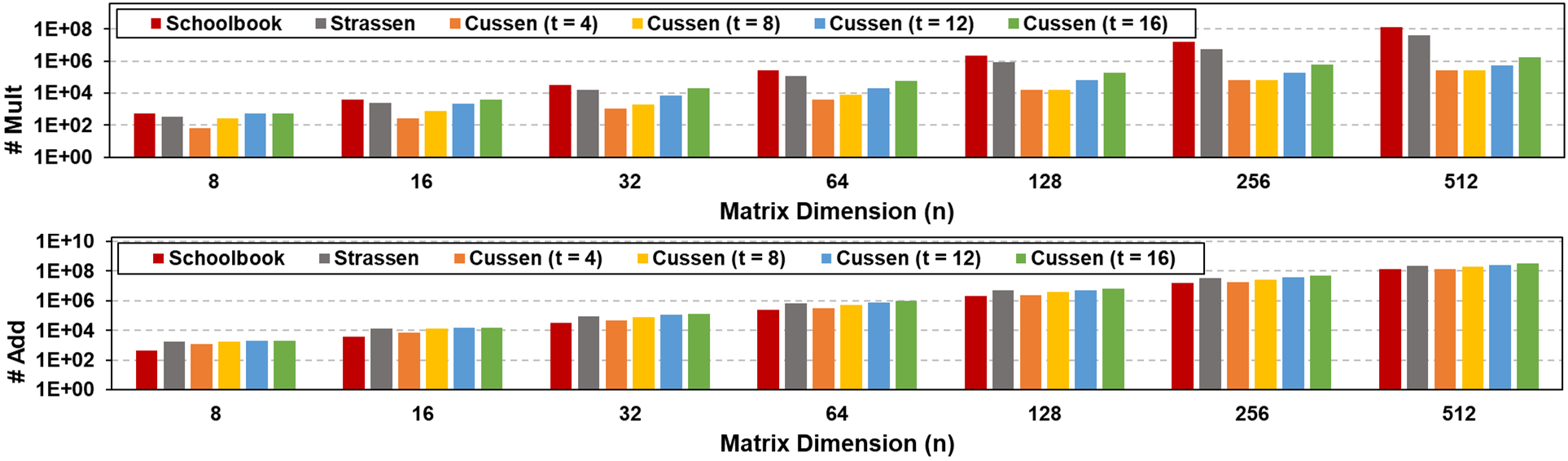}
\caption{Number of multiplications and additions required for plaintext matrix multiplication using schoolbook approach, Strassen's algorithm and Cussen's algorithm for random square matrices of dimension $n \in \{2^3, 2^4, \cdots, 2^9\}$ with element bit-widths $t \in \{4, 8, 12, 16\}$.}
\label{fig:cussen_plaintext_matrix_python}
\end{figure} 

To understand the benefits of Cussen's plaintext vector-scalar multiplication algorithm, we analyze the number of multiplications required after compression and the number of additions required for reconstruction. We consider random vectors of length $n \in \{2^3, 2^4, 2^5, 2^6, 2^7, 2^8, 2^9\}$ with element bit-widths $t \in \{4, 8, 12, 16\}$ and obtain the multiplication and addition counts averaged over 1000 random trials for each $(n, t)$ combination with 4 iterations of compression and reconstruction using our Python implementation of Cussen's algorithm. The results are shown in Figure \ref{fig:cussen_plaintext_vector_python} along with comparison with the schoolbook approach (which simply requires $n$ element-wise multiplications and zero additions).
We observe that Cussen's algorithm is able to significantly reduce the multiplication count (up to 2 orders of magnitude) in case of large vector sizes and relatively small element bit-widths by taking repeated differences and eliminating possible duplicates in the compression phase. Of course, this comes at the cost of a large number of additions in the reconstruction phase, which are not required in the schoolbook approach.
Data used to plot Figure \ref{fig:cussen_plaintext_vector_python} are provided in Tables \ref{table:appendix:cussen_plaintext_vector_python_mul} and \ref{table:appendix:cussen_plaintext_vector_python_add} in the Appendix.

This can be generalized to matrix-matrix multiplication by applying the above methodology to each column of matrix $\boldsymbol{A}$ and then multiplying with elements of matrix $\boldsymbol{B}$. Note that the compression phase can be amortized across all the columns of an outer product, while the reconstruction phase must be performed for each column separately after multiplications with the compressed version.
To understand the benefits of Cussen's algorithm compared to matrix multiplication using schoolbook method and Strassen's algorithm, we analyze their computational requirements using Python simulations.
For simplicity of presentation, we consider square matrices ($m = n = l$) with dimensions $n \in \{2^3, 2^4, 2^5, 2^6, 2^7, 2^8, 2^9\}$ and element bit-widths $t \in \{4, 8, 12, 16\}$.
For Cussen's algorithm, we extrapolate from the vector-scalar multiplication analysis discussed above to the matrix multiplication case.
The results are summarized in Figure \ref{fig:cussen_plaintext_matrix_python} and we observe that Cussen's algorithm is able to significantly reduce the number of multiplications (up to 2 orders of magnitude) while increasing the number of additions in case of large matrices with relatively small element bit-widths.
However, an interesting observation is that the number of additions required in both Strassen's and Cussen's algorithms are still comparable (both are larger than the $n^2 (n-1)$ additions required in schoolbook approach).
Data used to plot Figure \ref{fig:cussen_plaintext_matrix_python} are provided in Tables \ref{table:appendix:cussen_plaintext_matrix_python_mul} and \ref{table:appendix:cussen_plaintext_matrix_python_add} in the Appendix.

\subsection{Proposed Approach to PC-MM from Unpacked AHE}
\label{subsec:design:proposed}

Although proposed with the motivation to improve hardware efficiency, Cussen's algorithm has found little adoption so far in practical implementation, e.g., in machine learning acceleration. This is possibly due to the fact that computer architecture and semiconductor technology innovations have already made multiplication circuits quite efficient, and data movement between memory and computational units has become the major performance bottleneck in such applications rather than the computations themselves \cite{horowitz_isscc_2014}.

In this work, we demonstrate a suitable application of Cussen's algorithm in a very different context of privacy-preserving computation which can truly exploit its advantage.
We extend the plaintext compression-reconstruction technique to the encrypted setting for plaintext-ciphertext matrix multiplication (\texttt{PC-MM}) with unpacked additively homomorphic encryption (\texttt{AHE}).
In particular, we focus on the elliptic curve ElGamal (EC-ElGamal) encryption scheme discussed in Sections \ref{subsec:background:ecc} and \ref{subsec:background:elgamal}.
Here, our key observation is that multiplication of a ciphertext by a plaintext scalar requires two ECSM computations while adding two ciphertexts requires two point additions, since an EC-ElGamal ciphertext is a tuple of points as explained in Section \ref{subsec:background:elgamal}.
Now, an ECSM computation with a $t$-bit scalar using Algorithm \ref{algo:ecsm} requires $t$ point doubling and $t$ point addition operations. While the relative costs of implementing point doubling and point addition using prime field arithmetic are different for different elliptic curves \cite{bernsteinlange_efd}, we assume they are approximately the same for simplicity of analysis. Then, multiplication of a ciphertext by a plaintext scalar is $\approx 2t$ times more expensive than addition of two ciphertexts. Therefore, we have a suitable setting where ``multiplications'' (plaintext-ciphertext) are significantly more expensive than ``additions'' (ciphertext-ciphertext) for reasonably large scalars.

With this motivation, we propose an efficient approach to compute \texttt{PC-MM} with EC-ElGamal-based unpacked \texttt{AHE} by applying Cussen's algorithm. This is summarized in Algorithm \ref{algo:pcmm} with an $m \times n$ plaintext matrix $\boldsymbol{A} = [a_{ik}]_{m \times n}$ and $nl$ EC-ElGamal ciphertexts corresponding to an $n \times l$ plaintext matrix $\boldsymbol{B} = [b_{kj}]_{n \times l}$ as the inputs, and $ml$ EC-ElGamal ciphertexts corresponding to an $m \times l$ plaintext matrix $\boldsymbol{C} = \boldsymbol{A} \times \boldsymbol{B} = [c_{ij}]_{m \times l}$ as the outputs. Here, step 3 compresses the $k$-th column of $\boldsymbol{A}$, while steps 5 and 6 reconstruct its products with the ciphertexts corresponding to the $(k,j)$-th elements of $\boldsymbol{B}$. Step 8 does column-wise summation of the encrypted outer products to obtain the final ciphertexts corresponding to $\boldsymbol{C}$.
The compression in step 3 gets amortized across all $l$ columns of each of the $n$ outer products. Therefore, the core of our proposed \texttt{PC-MM} is an application of Cussen's efficient scalar-vector multiplication algorithm in the encrypted setting, and the same is then repeated $nl$ times to obtain the final plaintext-ciphertext matrix product.
It is important to note that this method always performs exact reconstruction of the final ciphertext matrix / vector, and there is no loss in accuracy compared to the plaintext result. Of course, the plaintext matrix / vector elements need to be integers as implicitly defined by the underlying \texttt{AHE} scheme and its message space $\cal{M}$. Possible solutions for handling real numbers are discussed in Section \ref{subsec:implementation:application}.

\begin{figure}[!t]
\centering
\includegraphics[width=0.98\textwidth]{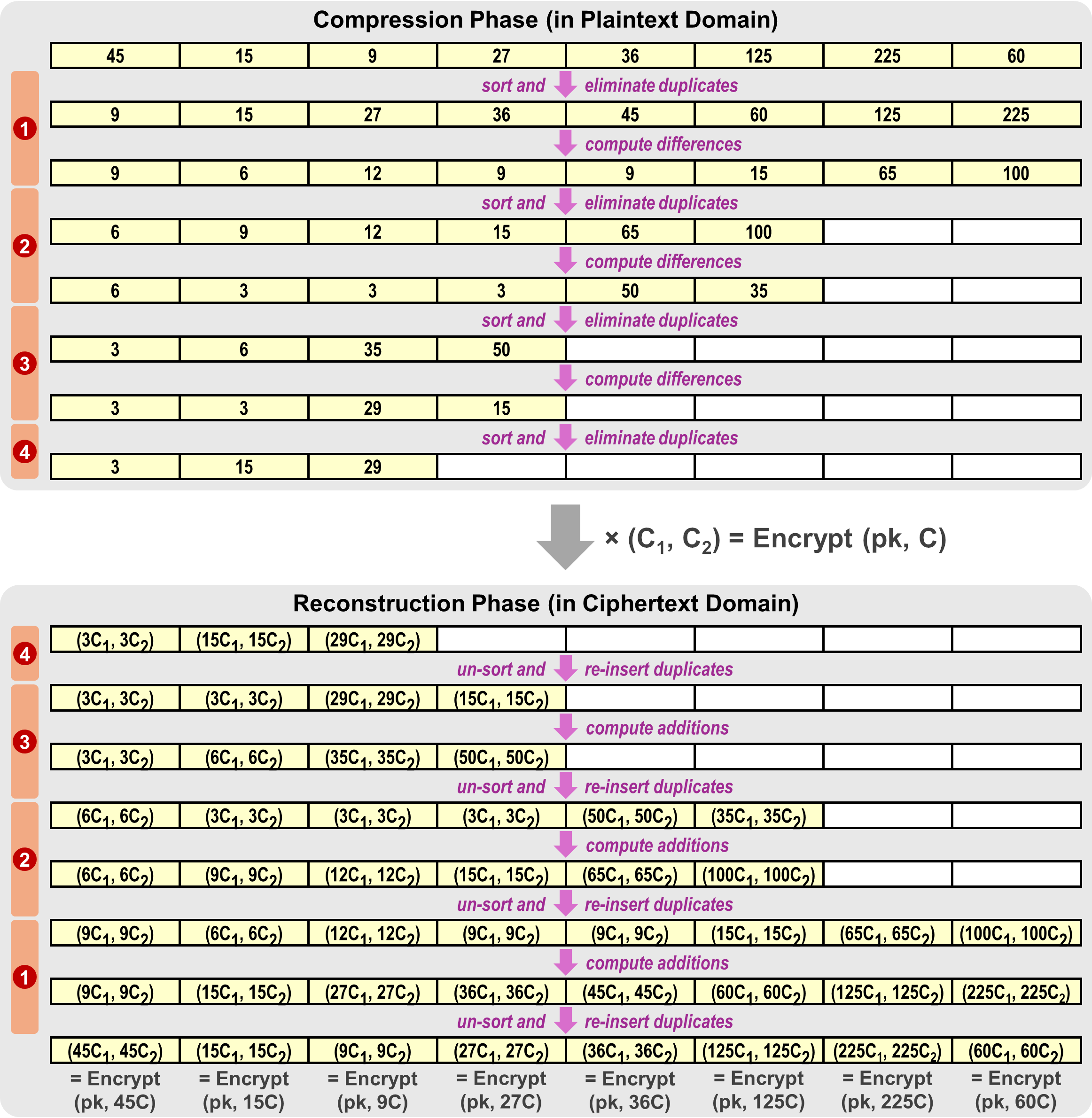}
\caption{Toy example showing efficient multiplication of plaintext vector and EC-ElGamal ciphertext corresponding to encrypted scalar using proposed approach with four iterations of Cussen's compression-reconstruction algorithm.}
\label{fig:cussen_algo_encrypted}
\end{figure}

Figure \ref{fig:cussen_algo_encrypted} shows how the toy example of plaintext vector-scalar multiplication from Figure \ref{fig:cussen_algo} can be extended to the encrypted setting. This example multiplies a plaintext vector of length $n = 8$ with 8-bit elements and an EC-ElGamal ciphertext $(C_1, C_2)$ corresponding to plaintext scalar $C$.
The compression phase remains exactly the same as in the plaintext setting, while the reconstruction phase in the encrypted setting now involves elliptic curve scalar multiplication (ECSM) operations and elliptic curve point additions.
Compared to 16 ECSM operations in the schoolbook approach, the example in Figure \ref{fig:cussen_algo_encrypted} requires 6 ECSM operations and 30 point additions. For 8-bit scalars, the cost of 1 ECSM operation is approximately equivalent to 16 point additions, as discussed earlier. Therefore, our proposed approach using Cussen's compression-reconstruction algorithm provides a clear advantage by saving computation cost equivalent to 130 point additions in this toy example. The same can be extended to the case of \texttt{PC-MM} following Algorithm \ref{algo:pcmm}.

\begin{algorithm}[!t]
\caption{Proposed efficient \texttt{PC-MM} with EC-ElGamal-based unpacked \texttt{AHE} using Cussen's compression-reconstruction algorithm}
\label{algo:pcmm}
\begin{algorithmic}[1]
\REQUIRE plaintext matrix $\boldsymbol{A} = [a_{ik}]_{m \times n}$ and $nl$ EC-ElGamal ciphertexts $(B_1^{(kj)}, B_2^{(kj)}) = \texttt{Encrypt} (pk, b_{kj})$ corresponding to plaintext matrix $\boldsymbol{B} = [b_{kj}]_{n \times l}$
\ENSURE $ml$ EC-ElGamal ciphertexts $(C_1^{(ij)}, C_2^{(ij)}) = \texttt{Encrypt} (pk, c_{ij})$ corresponding to plaintext matrix $\boldsymbol{C} = \boldsymbol{A} \times \boldsymbol{B} = [c_{ij}]_{m \times l}$ where $c_{ij} = \sum_{k = 1}^{n} a_{ik}b_{kj}$
\STATE $(C_1^{(ij)}, C_2^{(ij)}) \leftarrow (\infty, \infty)$ $\forall$ $1 \le i \le m$ and $1 \le j \le l$
\FOR{($k = 1$; $k \le n$; $k = k + 1$)}
\STATE Convert the $k$-th column $[a_{1k}, a_{2k}, \cdots, a_{mk}]^T$ of $\boldsymbol{A}$ to vector $[a'_{1k}, \cdots, a'_{m'k}]^T$ of length $m' \le m$ with multiple iterations of Cussen's compression algorithm
\FOR{($j = 1$; $j \le l$; $j = j + 1$)}
\STATE Multiply $[a'_{1k}, \cdots, a'_{m'k}]^T$ with $(B_1^{(kj)}, B_2^{(kj)})$ element-wise using $2m'$ ECSMs to get $[(a'_{1k}B_1^{(kj)}, a'_{1k}B_2^{(kj)}), \cdots, (a'_{m'k}B_1^{(kj)}, a'_{m'k}B_2^{(kj)})]^T$
\STATE Obtain $m$ ciphertexts corresponding to the $j$-th column of the $k$-th outer product as $[(a_{1k}B_1^{(kj)}, a_{1k}B_2^{(kj)}), (a_{2k}B_1^{(kj)}, a_{2k}B_2^{(kj)}), \cdots, (a_{mk}B_1^{(kj)}, a_{mk}B_2^{(kj)})]^T$ from $[(a'_{1k}B_1^{(kj)}, a'_{1k}B_2^{(kj)}), \cdots, (a'_{m'k}B_1^{(kj)}, a'_{m'k}B_2^{(kj)})]^T$ with multiple iterations of Cussen's reconstruction algorithm using elliptic curve point additions
\FOR{($i = 1$; $i \le m$; $i = i + 1$)}
\STATE $(C_1^{(ij)}, C_2^{(ij)}) \leftarrow (C_1^{(ij)} + a_{ik}B_1^{(kj)}, C_2^{(ij)} + a_{ik}B_2^{(kj)})$
\ENDFOR
\ENDFOR
\ENDFOR
\RETURN $(C_1^{(ij)}, C_2^{(ij)})$ $\forall$ $1 \le i \le m$ and $1 \le j \le l$
\end{algorithmic}
\end{algorithm}

\begin{figure}[!t]
\centering
\includegraphics[width=\textwidth]{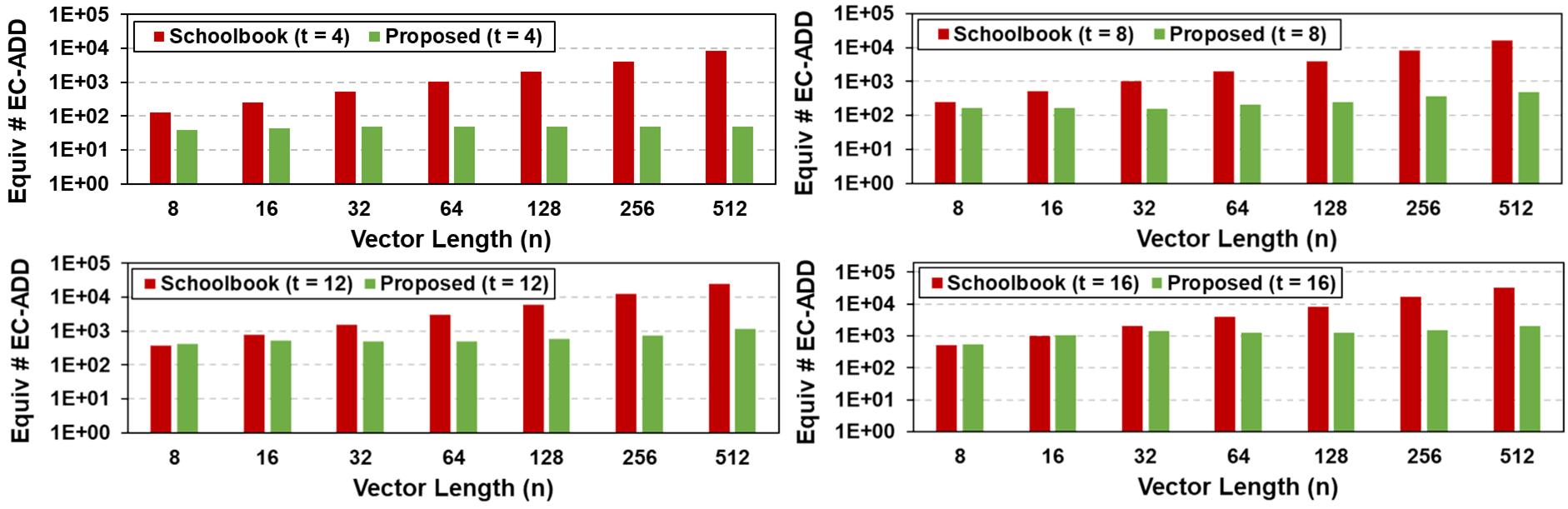}
\caption{Equivalent number of elliptic curve point additions required for plaintext vector and ciphertext scalar multiplication using schoolbook approach and proposed approach based on Cussen's algorithm for random vectors of length $n \in \{2^3, 2^4, \cdots, 2^9\}$ with element bit-widths $t \in \{4, 8, 12, 16\}$.}
\label{fig:cussen_ciphertext_vector_python}
\end{figure}

Again, we first analyze the benefits of Cussen's compression-reconstruction algorithm in the context of multiplying plaintext vectors with ciphertext scalars.
We consider random vectors of length $n \in \{2^3, 2^4, 2^5, 2^6, 2^7, 2^8, 2^9\}$ with element bit-widths $t \in \{4, 8, 12, 16\}$ and obtain the operation counts averaged over 1000 random trials for each $(n, t)$ combination with 4 iterations of compression and reconstruction using our Python implementation. The results are shown in Figure \ref{fig:cussen_ciphertext_vector_python} along with comparison with the schoolbook approach (which simply requires $2n$ ECSMs) in terms of the equivalent number of elliptic curve point additions (where we consider point doublings to be approximately equivalent to point additions).
We observe that by applying Cussen's algorithm, we are able to significantly reduce the equivalent point addition count (up to 2 orders of magnitude), primarily due to the compression phase, in case of large vector sizes and relatively small element bit-widths, even after accounting for extra point additions required in the reconstruction phase. 
Data used to plot Figure \ref{fig:cussen_ciphertext_vector_python} are provided in Table \ref{table:appendix:cussen_ciphertext_vector_python} in the Appendix.

\begin{figure}[!t]
\centering
\includegraphics[width=\textwidth]{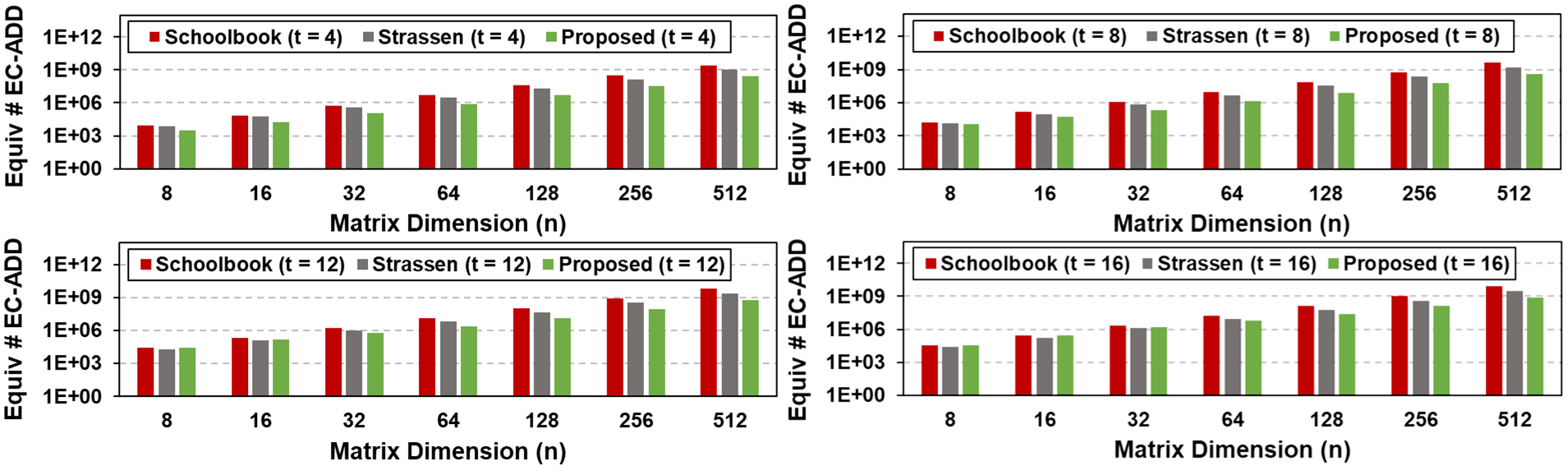}
\caption{Equivalent number of elliptic curve point additions required for plaintext-ciphertext matrix multiplication \texttt{PC-MM} using schoolbook approach, Strassen's algorithm and proposed approach based on Cussen's algorithm for random square matrices of dimension $n \in \{2^3, 2^4, \cdots, 2^9\}$ with element bit-widths $t \in \{4, 8, 12, 16\}$.}
\label{fig:cussen_ciphertext_matrix_python}
\end{figure}

This analysis confirms our claim that it is indeed possible to reap the benefits of Cussen's compression-reconstruction algorithm in this encrypted setting.
So, we finally compare the equivalent number of elliptic curve point additions required for plaintext-ciphertext matrix multiplication \texttt{PC-MM} using schoolbook approach, Strassen's algorithm and our proposed approach (Algorithm \ref{algo:pcmm}) based on Cussen's algorithm.
For simplicity of presentation, we again consider square matrices ($m = n = l$) with dimensions $n \in \{2^3, 2^4, 2^5, 2^6, 2^7, 2^8, 2^9\}$ and element bit-widths $t \in \{4, 8, 12, 16\}$, and use Python simulations to obtain the results shown in Figure \ref{fig:cussen_ciphertext_matrix_python}.
For our proposed approach using Cussen's algorithm, we extrapolate from the plaintext vector and ciphertext scalar multiplication analysis discussed above to the plaintext-ciphertext matrix multiplication case.
Clearly, we are able to achieve significant reduction in the equivalent point addition count (up to an order of magnitude) for large matrices with relatively small element bit-widths, improving even further beyond what is possible using Strassen's algorithm. 
Data used to plot Figure \ref{fig:cussen_ciphertext_matrix_python} are provided in Table \ref{table:appendix:cussen_ciphertext_matrix_python} in the Appendix.
\section{Implementation and Analysis}
\label{sec:implementation}

\subsection{Software Implementation and Experimental Setup}
\label{subsec:implementation:software}

We experimentally validate our proposed approach for fast \texttt{PC-MM} with Cussen's compression-reconstruction algorithm using a software implementation of the elliptic curve ElGamal additively homomorphic public key encryption scheme and its extension to plaintext-ciphertext matrix multiplication evaluations.
Previous work have mostly evaluated their techniques on high-performance desktop or server-scale processors \cite{halevi_helib_2014, phong_deep_2017, juvekar_gazelle_2018, froelicher_federated_2023, liu_verifiable_2023}, while few have presented implementations on embedded micro-controllers and IoT platforms \cite{reddy_wfiot_2022, ahsan_ants_2022, banerjee_globecom_2023}.
It is generally believed that homomorphic computations on encrypted data are prohibitively expensive in the IoT due to their high computational overheads \cite{reddy_wfiot_2022}.
In this work, we attempt to overcome this barrier and demonstrate fast \texttt{PC-MM} from EC-ElGamal-based unpacked \texttt{AHE} on a Raspberry Pi IoT platform.
Our software implementation is based on the open-source MIRACL cryptographic library with efficient and IoT-friendly realizations of elliptic curve operations \cite{scott_ecciot_2020}: \url{https://github.com/miracl/core}.
The code we have used for our experiments is available online in the following repository along with usage instructions: \url{https://github.com/sinesyslab/fast_pcmm_ahe_cic25}.

We use the NIST standard P-256 elliptic curve $E(\Fp): y^2 = x^3 - 3x + b$ (mod $p$) defined over a 256-bit prime field (with $p = 2^{256} - 2^{224} + 2^{192} + 2^{96} - 1$) and having a 256-bit prime order $q$.
We use the MIRACL function \texttt{ECP\_NIST256\_pinmul} to compute ECSM with any point $P \in E(\Fp)$ and any small $t$-bit scalar $k \ll q$ using the Montgomery ladder method from Algorithm \ref{algo:ecsm}. The MIRACL implementation is constant-time and resilient against certain classes of simple timing and power analysis side-channel attacks \cite{fan_ecc_2010}. In this work, we primarily focus on demonstrating and profiling the fast \texttt{PC-MM} implementation, and side-channel analysis is out of the scope.
For elliptic curve point additions, we use the function \texttt{ECP\_NIST256\_add} available in the MIRACL library.

\begin{figure}[!t]
\centering
\includegraphics[width=\textwidth]{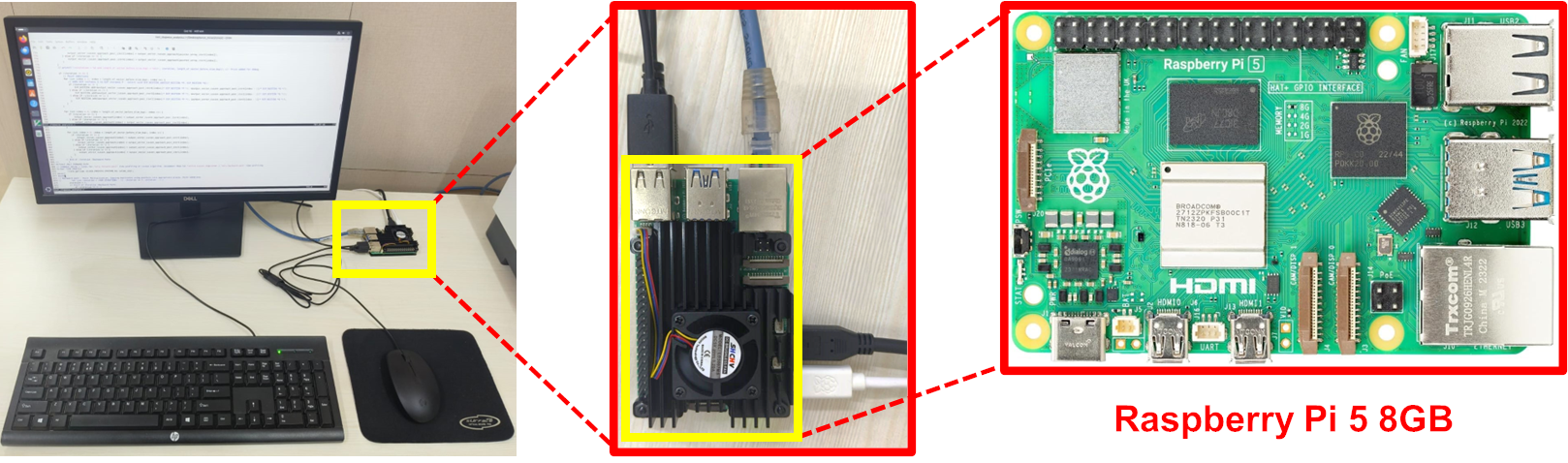}
\caption{Experimental setup used for evaluating and profiling proposed fast \texttt{PC-MM}.}
\label{fig:setup}
\end{figure}

We choose the Raspberry Pi 5 single board computer (with a 2.4~GHz quad-core 64-bit ARM Cortex-A76-based Broadcom BCM2712 system-on-chip, an 8~GB LPDDR4X-4267 SDRAM off-chip memory and a 128~GB microSDXC A2/V30/U3 UHS-I persistent storage) as our experimental platform as it is generally considered a good representative of an edge computing device.
We use the Ubuntu 24.04 LTS Linux operating system and all C programs are compiled using the GCC compiler version 13.2.0-23ubuntu4 with the -O3 optimization flag.
Our experimental setup is shown in Figure \ref{fig:setup}.
Note that our proposed approach for fast \texttt{PC-MM} is also expected to work well for any other choice of elliptic curve, software library, operating system and evaluation platform.

\subsection{Measurement Results and Performance Analysis}
\label{subsec:implementation:analysis}

We present the experimentally measured performance results of our proposed approach (as outlined in Algorithm \ref{algo:pcmm} including both compression and reconstruction phases) to \texttt{PC-MM} using EC-ElGamal-based unpacked \texttt{AHE} explained in Section \ref{subsec:design:proposed} with the software implementation and evaluation setup described in Section \ref{subsec:implementation:software}.

\begin{figure}[!t]
\centering
\includegraphics[width=\textwidth]{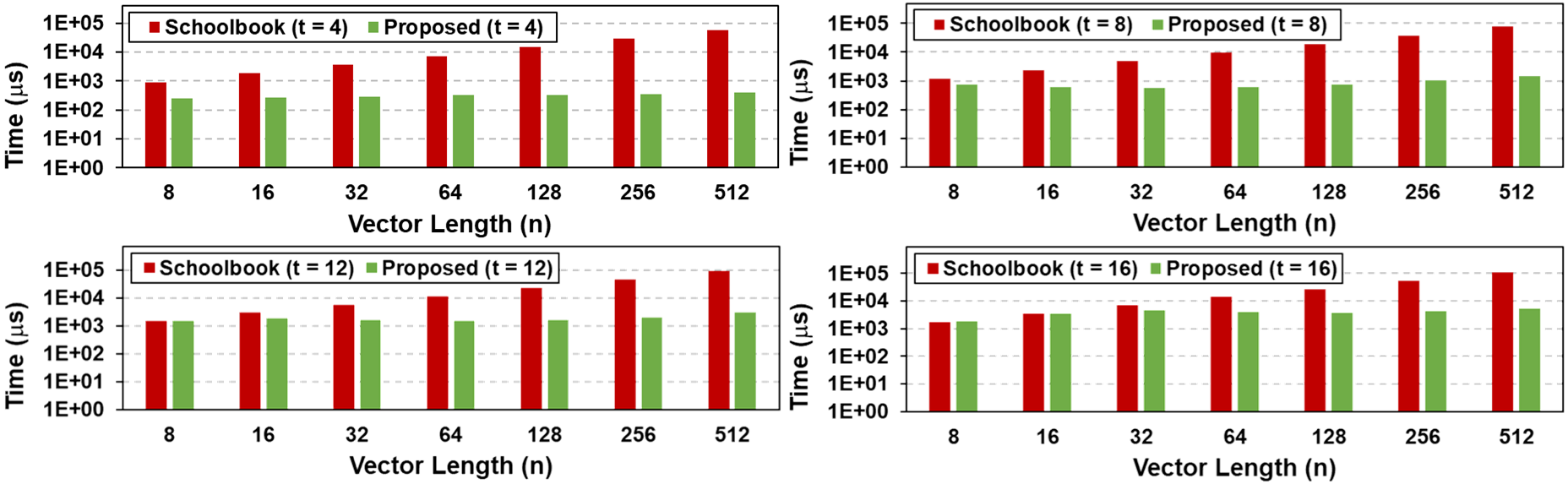}
\caption{Time taken for plaintext vector and ciphertext scalar multiplication using schoolbook approach and proposed approach based on Cussen's algorithm for random vectors of length $n \in \{2^3, 2^4, \cdots, 2^9\}$ with element bit-widths $t \in \{4, 8, 12, 16\}$.}
\label{fig:cussen_ciphertext_vector_rpi}
\end{figure}

\begin{figure}[!t]
\centering
\includegraphics[width=0.9\textwidth]{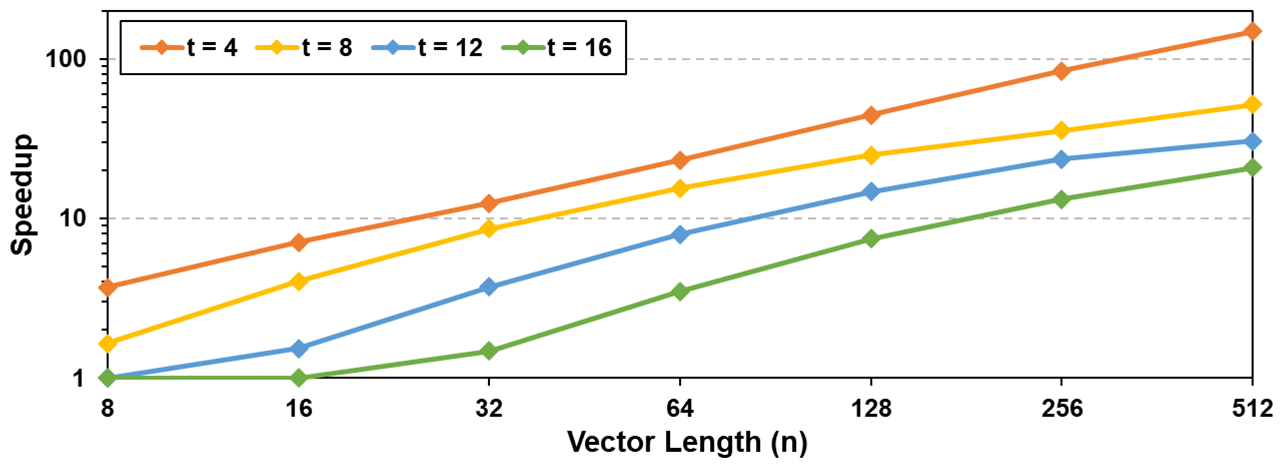}
\caption{Speedup observed for plaintext vector and ciphertext scalar multiplication using proposed approach based on Cussen's algorithm compared to schoolbook approach for random vectors of length $n \in \{2^3, 2^4, \cdots, 2^9\}$ with element bit-widths $t \in \{4, 8, 12, 16\}$.}
\label{fig:speedup_vector}
\end{figure}

First, we measure the time required for plaintext vector and ciphertext scalar multiplication using the schoolbook approach and our proposed approach using Cussen's algorithm (4 iterations of compression and reconstruction). Consistent with the analysis in Section \ref{subsec:design:proposed}, we consider random vectors of length $n \in \{2^3, 2^4, 2^5, 2^6, 2^7, 2^8, 2^9\}$ with element bit-widths $t \in \{4, 8, 12, 16\}$, and obtain the execution times averaged over 100 random trials for each $(n, t)$ combination using our software implementation on Raspberry Pi 5.
The results are shown and compared in Figure \ref{fig:cussen_ciphertext_vector_rpi}.
Data used to plot Figure \ref{fig:cussen_ciphertext_vector_rpi} are provided in Table \ref{table:appendix:cussen_ciphertext_vector_rpi} in the Appendix.
For better visualization, we also present the speedup compared to the schoolbook approach in Figure \ref{fig:speedup_vector}.
We observe speedups very similar to our Python simulations discussed in Section \ref{subsec:design:proposed} and presented in Figure \ref{fig:cussen_ciphertext_vector_python}. Note that the speedup is > 1 for all $(n, t)$ combinations except $(n, t) \in \{(8, 12), (8, 16), (16, 16)\}$. This can be easily explained by the fact that the probability of encountering duplicates in such short random vectors with large element bit-widths is negligible, thus making the compression phase of Cussen's algorithm ineffective. On the other hand, for large vectors with relatively small element bit-widths, the speedups are very large, e.g., $148\times$ for $(n = 512, t = 4)$, $84\times$ for $(n = 256, t = 4)$, $52\times$ for $(n = 512, t = 8)$, $44\times$ for $(n = 128, t = 4)$, $36\times$ for $(n = 256, t = 8)$, $31\times$ for $(n = 512, t = 12)$, $25\times$ for $(n = 128, t = 8)$, $24\times$ for $(n = 256, t = 12)$, $23\times$ for $(n = 64, t = 4)$ and $21\times$ for $(n = 512, t = 16)$.
Data used to plot Figure \ref{fig:speedup_vector} are provided in Table \ref{table:appendix:speedup_vector_rpi} in the Appendix.

\begin{figure}[!t]
\centering
\includegraphics[width=\textwidth]{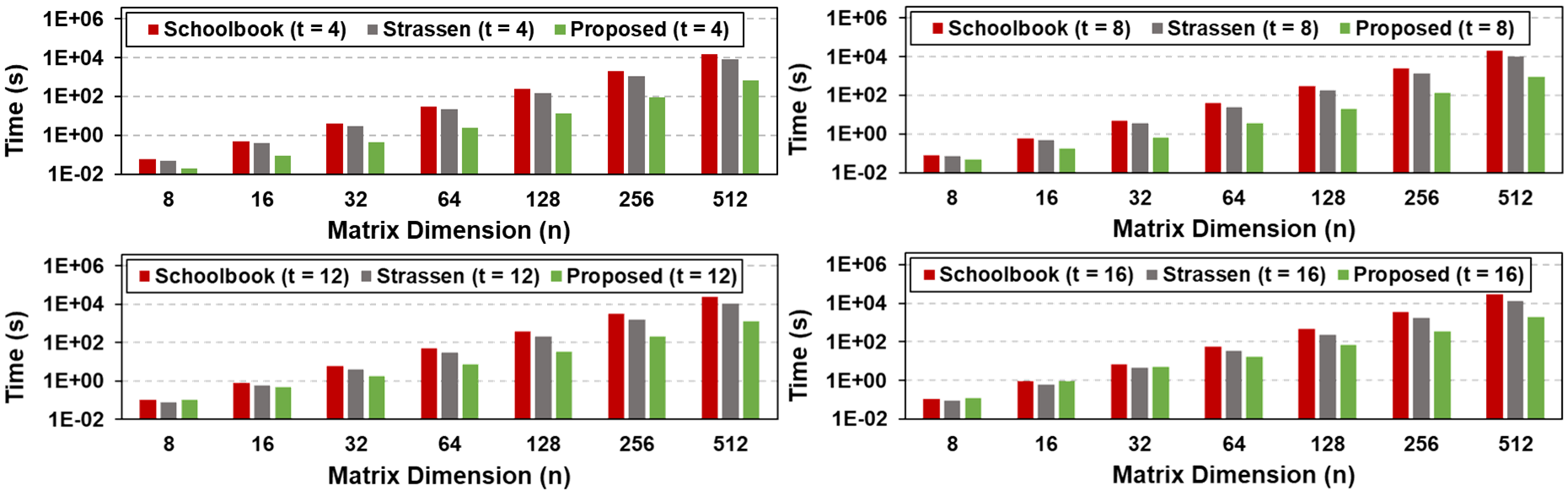}
\caption{Time taken for plaintext-ciphertext matrix multiplication \texttt{PC-MM} using schoolbook approach, Strassen's algorithm and proposed approach based on Cussen's algorithm for random square matrices of dimension $n \in \{2^3, 2^4, \cdots, 2^9\}$ with element bit-widths $t \in \{4, 8, 12, 16\}$.}
\label{fig:cussen_ciphertext_matrix_rpi}
\end{figure}

\begin{figure}[!t]
\centering
\includegraphics[width=\textwidth]{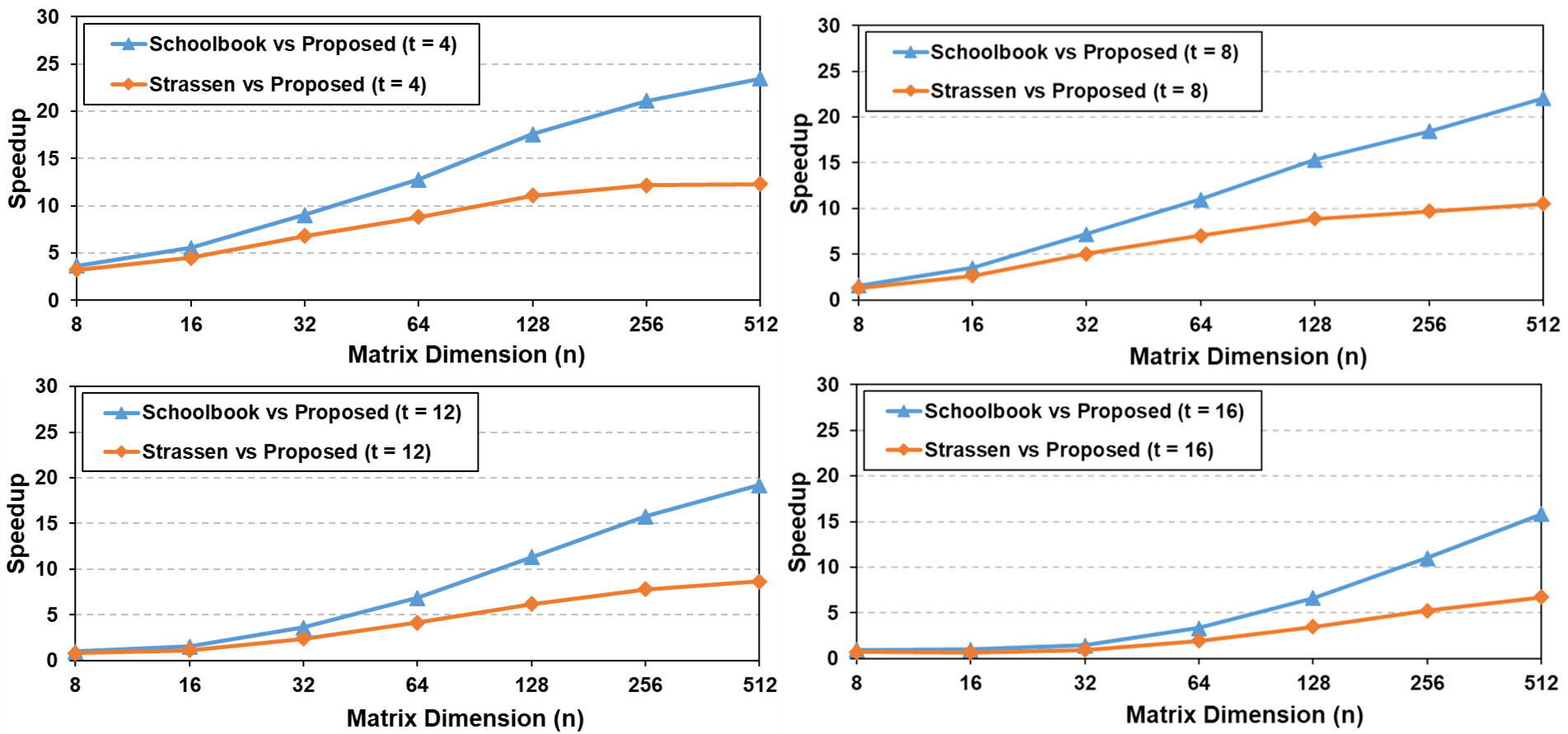}
\caption{Speedups observed for plaintext-ciphertext matrix multiplication \texttt{PC-MM} using proposed approach based on Cussen's algorithm compared to schoolbook approach and Strassen's algorithm for random square matrices of dimension $n \in \{2^3, 2^4, \cdots, 2^9\}$ with element bit-widths $t \in \{4, 8, 12, 16\}$.}
\label{fig:speedup_matrix}
\end{figure}

Finally, we measure the time required for plaintext-ciphertext matrix multiplication \texttt{PC-MM} using schoolbook approach, Strassen's algorithm and our proposed approach based on Cussen's algorithm (4 iterations of compression and reconstruction). Consistent with the analysis in Section \ref{subsec:design:proposed}, we consider square matrices ($m = n = l$) with dimensions $n \in \{2^3, 2^4, 2^5, 2^6, 2^7, 2^8, 2^9\}$ and element bit-widths $t \in \{4, 8, 12, 16\}$, and obtain the execution times for each $(n, t)$ combination using our software implementation on Raspberry Pi 5.
The results are shown and compared in Figure \ref{fig:cussen_ciphertext_matrix_rpi}.
Data used to plot Figure \ref{fig:cussen_ciphertext_matrix_rpi} are provided in Table \ref{table:appendix:cussen_ciphertext_matrix_rpi} in the Appendix.
For better visualization, we also present the speedups compared to the schoolbook approach and Strassen's algorithm in Figure \ref{fig:speedup_matrix}.
Again, we observe speedups similar to our Python simulations discussed in Section \ref{subsec:design:proposed} and presented in Figure \ref{fig:cussen_ciphertext_matrix_python}. Note that the speedup is > 1 for all $(n, t)$ combinations except $(n, t) \in \{(8, 12), (8, 16), (16, 16)\}$ due to the same reason explained earlier. As anticipated, for large matrices with relatively small element bit-widths, the speedups are large even compared to Strassen's algorithm, e.g., $12\times$ for $(n, t) \in \{(512, 4), (256, 4)\}$, $11\times$ for $(n, t) \in \{(512, 8), (128, 4)\}$ and $10\times$ for $(n = 256, t = 8)$.
Data used to plot Figure \ref{fig:speedup_matrix} are provided in Table \ref{table:appendix:speedup_matrix_rpi} in the Appendix.
To emphasize the significance of these speedups, we give a concrete example.
The \texttt{PC-MM} computation for $(n = 512, t = 8)$ with our MIRACL-based software implementation on Raspberry Pi 5 requires $\approx 5.72$ hours using the schoolbook approach and $\approx 2.73$ hours using Strassen's algorithm, while it is executed in just $\approx 15.6$ minutes using our proposed approach with Cussen's compression-reconstruction algorithm in the encrypted setting.
Our experimental results and performance analyses confirm that we have indeed been successful in lowering the computation barrier for \texttt{PC-MM} from unpacked \texttt{AHE} in the context of IoT, embedded systems and edge computing platforms.
While the execution times will be much faster in case of high-performance server-scale or desktop micro-processors, the same order of speedup is expected with our proposed approach.
Of course, the execution times across all three approaches can also be further reduced using dedicated hardware accelerators for elliptic curve cryptography \cite{quisquater_jsys_2007, rashidi_survey_2017, ifrim_trets_2024}.

\subsection{Applications and Extensions}
\label{subsec:implementation:application}

\subsubsection{Applications}

Plaintext-Ciphertext Matrix Multiplication (\texttt{PC-MM}) is an indispensable tool in privacy-preserving computations such as machine learning and signal processing in the encrypted setting (with plaintext weights and ciphertext data).
This is applicable not only in the cloud environment but also in IoT networks where \texttt{PC-MM} can be used by edge computing systems to extract useful information from sensor nodes without revealing sensitive data, e.g., encrypted sensor data classification and secure wireless fingerprint-based indoor localization \cite{li_infocom_2014, jarvinen_infocom_2018, ryffel_ipe_2019, ahsan_ants_2022, banerjee_globecom_2023}.

Matrices used to represent model parameters, such as weights, in machine learning are often sparse. In case of sparse plaintext matrices, the column vectors can be further compressed leading to faster \texttt{PC-MM} using our proposed approach.
Practical applications often need to deal with rectangular matrices, which can also be handled easily using our proposed approach by simply setting the appropriate parameters $m$, $n$ and $l$ in Algorithm \ref{algo:pcmm}.
Finally, computations such as 1-D / 2-D linear and circular convolutions can be represented as matrix multiplications with Toeplitz and circulant matrices \cite{nikou_convolution_2014, salehi_convolution_2022} and our proposed fast \texttt{PC-MM} approach can be used in such applications as well.

\subsubsection{Handling Real Numbers}

Our proposed \texttt{PC-MM} approach described in Sections \ref{subsec:design:proposed} and \ref{subsec:implementation:analysis} from the EC-ElGamal unpacked \texttt{AHE} scheme implicitly requires the matrix elements to be integers.
This feature is common with most other well-known homomorphic encryption schemes, both PHE and FHE (notable exceptions include the CKKS lattice-based FHE scheme \cite{ckks_fhe_2017} which natively supports approximate arithmetic with real numbers in the encrypted domain).
Similar to other schemes supporting encrypted integer computations, our proposed \texttt{PC-MM} can be extended to handle real numbers by encoding them as integers, e.g., by scaling and rounding. In other words, matrices whose elements are real floating-point numbers need to be converted to fixed-point representation to be suitable for the proposed \texttt{PC-MM}.
The impact of such encoding on accuracy is exactly the same in both plaintext and ciphertext as the \texttt{PC-MM} does not introduce any additional errors during compression and reconstruction.
Note that the original Cussen's algorithm for plaintext matrix multiplication, which has inspired our proposed \texttt{PC-MM}, is also primarily focused on integer arithmetic.
Possible optimizations to better handle real numbers will be explored in future work.

\subsubsection{Extension to Paillier Encryption}

Although we have demonstrated our proposed approach to fast \texttt{PC-MM} from EC-ElGamal-based unpacked \texttt{AHE}, it can be easily extended to the additively homomorphic Paillier encryption scheme \cite{paillier_pke_1999}.
As discussed in Section \ref{subsec:background:paillier}, in the context of Paillier-based unpacked \texttt{AHE}, multiplication of ciphertexts leads to addition of their underlying plaintexts and exponentiating a ciphertext to a plaintext scalar power leads to multiplication of its underlying plaintext with the scalar.
Therefore, for the Paillier \texttt{AHE} scheme,  considering an $m \times n$ plaintext matrix $\boldsymbol{A} = [a_{ik}]_{m \times n}$ and $nl$ ciphertexts $\texttt{Encrypt} (pk, b_{kj})$ corresponding to an $n \times l$ plaintext matrix $\boldsymbol{B} = [b_{kj}]_{n \times l}$, the $ml$ ciphertexts corresponding to the $m \times l$ plaintext matrix $\boldsymbol{A} \times \boldsymbol{B} = \boldsymbol{C} = [c_{ij}]_{m \times l}$ can be expressed as follows using the column-and-row outer product technique from Section \ref{subsec:background:matmul}:
\begin{align*}
\begin{bmatrix}
\texttt{Encrypt} (pk, c_{11}) & \texttt{Encrypt} (pk, c_{12}) & \cdots & \texttt{Encrypt} (pk, c_{1l}) \\
\texttt{Encrypt} (pk, c_{21}) & \texttt{Encrypt} (pk, c_{22}) & \cdots & \texttt{Encrypt} (pk, c_{2l}) \\
\vdots & \vdots & \ddots & \vdots \\
\texttt{Encrypt} (pk, c_{m1}) & \texttt{Encrypt} (pk, c_{m2}) & \cdots & \texttt{Encrypt} (pk, c_{ml}) \\
\end{bmatrix} \nonumber \\
= 
\begin{bmatrix}
\prod\limits_{k = 1}^{n}\texttt{Encrypt} (pk, b_{k1}) ^ {a_{1k}} & \prod\limits_{k = 1}^{n}\texttt{Encrypt} (pk, b_{k2}) ^ {a_{1k}} & \cdots & \prod\limits_{k = 1}^{n}\texttt{Encrypt} (pk, b_{kl}) ^ {a_{1k}} \\
\prod\limits_{k = 1}^{n}\texttt{Encrypt} (pk, b_{k1}) ^ {a_{2k}} & \prod\limits_{k = 1}^{n}\texttt{Encrypt} (pk, b_{k2}) ^ {a_{2k}} & \cdots & \prod\limits_{k = 1}^{n}\texttt{Encrypt} (pk, b_{kl}) ^ {a_{2k}} \\
\vdots & \vdots & \ddots & \vdots \\
\prod\limits_{k = 1}^{n}\texttt{Encrypt} (pk, b_{k1}) ^ {a_{mk}} & \prod\limits_{k = 1}^{n}\texttt{Encrypt} (pk, b_{k2}) ^ {a_{mk}} & \cdots & \prod\limits_{k = 1}^{n}\texttt{Encrypt} (pk, b_{kl}) ^ {a_{mk}} \\
\end{bmatrix}
\end{align*}
which requires total $mln$ ciphertext exponentiations to plaintext powers and $ml(n-1)$ ciphertext-ciphertext multiplications.

\begin{algorithm}[!t]
\caption{Exponentiation using Montgomery powering ladder \cite{joye_montgomeryladder_2002}}
\label{algo:modexp}
\begin{algorithmic}[1]
\REQUIRE $k = (k_{t-1}, \cdots, k_1, k_0)_2$ and $g$
\ENSURE $g^k$
\STATE $r_0 \leftarrow 1$, $r_1 \leftarrow g$
\FOR{($i = t-1$; $i \ge 0$; $i = i - 1$)}
\STATE $b \leftarrow k_i$
\STATE $r_{1-b} \leftarrow r_1 \cdot r_0$, $r_b \leftarrow r_b^2$
\ENDFOR
\RETURN $r_0$
\end{algorithmic}
\end{algorithm}

Now, ciphertext exponentiations can be efficiently computed using the \textit{Montgomery powering ladder} (similar to the Montgomery ladder for ECSM from Algorithm \ref{algo:ecsm}), as shown in Algorithm \ref{algo:modexp}, where all arithmetic is performed modulo the appropriate Paillier modulus.
This algorithm requires $t$ modular squaring and $t$ modular multiplication operations for exponentiation to a $t$-bit scalar power. Again, assuming that the implementation costs of modular squaring and multiplication are similar, a ciphertext exponentation is $\approx 2t$ times more expensive than multiplication of ciphertexts.
In other words, multiplication of a plaintext under Paillier encryption with another plaintext scalar is significantly more expensive than addition of two plaintexts under Paillier encryption for reasonably large scalars.
Therefore, we again arrive at a setting exactly similar to Section \ref{subsec:design:proposed} and the same analysis for EC-ElGamal \texttt{AHE} applies directly to Paillier \texttt{AHE}, including the Python profiling results (except that elliptic curve scalar multiplications and point additions are replaced by modular exponentiations and multiplications respectively).
Implementation results will depend on the software library used for Paillier encryption and its underlying modular arithmetic. While execution times are expected to be significantly longer than EC-ElGamal due to the use of larger moduli, the speedups compared to schoolbook approach and Strassen's algorithm are expected to be quite similar.
\section{Conclusions and Future Work}
\label{sec:conclusion}

In this work, we have presented an efficient approach to plaintext-ciphertext matrix multiplication (\texttt{PC-MM}) from unpacked additively homomorphic encryption (\texttt{AHE}) by applying Cussen's compression-reconstruction algorithm for plaintext-plaintext matrix multiplication in the encrypted setting.
In particular, we have considered the elliptic curve ElGamal additively homomorphic public key encryption scheme where PC-MM requires elliptic curve scalar multiplications and elliptic curve point additions.
Our key observation is that elliptic curve scalar multiplications are significantly more expensive than elliptic curve point additions for reasonably large scalars.
Therefore, compressing the columns of the plaintext matrix can help reduce the number of elliptic curve scalar multiplications at the cost of increased number of elliptic curve point additions required to reconstruct the multiplied columns for outer product computation.
To understand the benefits of our proposed approach, we have first compared its computation cost with the schoolbook method and Strassen's algorithm using Python simulations with random matrices.
Finally, we experimentally validate our proposed technique using a software implementation with the open-source MIRACL cryptographic library on a Raspberry Pi 5 edge computing platform.
Our measurement results indicate up to an order of magnitude speedup with our proposed \texttt{PC-MM} compared to Strassen's algorithm in case of large matrices with relatively small element bit-widths.
Such large matrices with constrained elements are quite common in practical applications like machine learning and signal processing, thus making our fast \texttt{PC-MM} an excellent candidate for efficient privacy-preserving computation.

Possible future extensions of this work include evaluating the proposed \texttt{PC-MM} with other unpacked \texttt{AHE} schemes, efficient handling of real numbers, exploring the possibility of ciphertext packing in such traditionally unpacked schemes, and exploring the feasibility of applying Cussen's compression-reconstruction algorithm in the context of lattice-based FHE schemes supporting ciphertext packing and SIMD-style processing.

\section*{Acknowledgment}

This work was supported by the Prime Minister's Research Fellowship, Ministry of Education, Government of India.
The authors would like to thank the anonymous reviewers for their valuable comments, suggestions and constructive feedback.


\bibliography{references}



\section*{Appendix}

\appendix
\section{Detailed Profiling Results}
\label{appendix}

Here, we provide detailed tables containing our experimental profiling data used in the plots in Sections \ref{sec:design} and \ref{sec:implementation}.
First, we tabulate our Python implementation results which we have used to compare different algorithms in terms of: (1) the number of integer multiplications and additions required for plaintext vector-scalar multiplication as well as plaintext matrix multiplication, and (2) the equivalent number of elliptic curve point additions required for plaintext vector and ciphertext scalar multiplication as well as plaintext-ciphertext matrix multiplication.
Then, we tabulate the experimental results from our MIRACL-based software implementation measured on Raspberry Pi 5 to compare the above in terms of absolute execution time as well as relative speedup.

\subsection{Analysis with Python}

\begin{table}[!ht]
\centering
\renewcommand{\arraystretch}{1.25}
\scriptsize
\begin{tabular}{|c|c|c|c|c|c|} \hline      
 $n$ &Schoolbook & Cussen $(t = 4)$ & Cussen $(t = 8)$ & Cussen $(t = 12)$ & Cussen $(t = 16)$ \\ \hline 
 8&8 & 1 & 4 & 8 & 8 \\ \hline 
 16&16 & 1 & 3 & 9 & 15 \\ \hline 
 32&32 & 1 & 2 & 7 & 20 \\ \hline 
 64&64 & 1 & 2 & 5 & 15 \\ \hline 
 128&128 & 1 & 1 & 4 & 11 \\ \hline 
 256&256 & 1 & 1 & 3 & 9 \\ \hline 
 512&512 & 1 & 1 & 2 & 7 \\ \hline
\end{tabular}
\caption{Number of multiplications required for plaintext vector-scalar multiplication using schoolbook approach and Cussen's algorithm for random vectors of length $n \in \{2^3, 2^4, \cdots, 2^9\}$ with element bit-widths $t \in \{4, 8, 12, 16\}$ (data for Figure \ref{fig:cussen_plaintext_vector_python}).}
\label{table:appendix:cussen_plaintext_vector_python_mul}
\end{table}

\begin{table}[!ht]
\centering
\renewcommand{\arraystretch}{1.25}
\scriptsize
\begin{tabular}{|c|c|c|c|c|c|} \hline      
 $n$
&Schoolbook & Cussen $(t = 4)$ & Cussen $(t = 8)$ & Cussen $(t = 12)$ & Cussen $(t = 16)$ \\ \hline 
 8
&0 & 11 & 21 & 24 & 24 \\ \hline 
 16
&0 & 14 & 35 & 46 & 48 \\ \hline 
 32
&0 & 16 & 49 & 79 & 94 \\ \hline 
 64
&0 & 17 & 73 & 128 & 172 \\ \hline 
 128
&0 & 17 & 111 & 197 & 288 \\ \hline 
 256
&0 & 17 & 169 & 307 & 487 \\ \hline 
 512&0 & 17 & 225 & 523 & 781 \\ \hline
\end{tabular}
\caption{Number of additions required for plaintext vector-scalar multiplication using schoolbook approach and Cussen's algorithm for random vectors of length $n \in \{2^3, 2^4, \cdots, 2^9\}$ with element bit-widths $t \in \{4, 8, 12, 16\}$ (data for Figure \ref{fig:cussen_plaintext_vector_python}).}
\label{table:appendix:cussen_plaintext_vector_python_add}
\end{table}

\begin{table}[!ht]
\centering
\renewcommand{\arraystretch}{1.25}
\scriptsize
\begin{tabular}{|c|c|c|c|c|c|c|} \hline       
 $n$
&Schoolbook & Strassen & Cussen $(t = 4)$ & Cussen $(t = 8)$ & Cussen $(t = 12)$ & Cussen $(t = 16)$ \\ \hline 
 8
&512 & 343 & 64 & 256 & 512 & 512 \\ \hline 
 16
&4096 & 2401 & 256 & 768 & 2304 & 3840 \\ \hline 
 32
&32768 & 16807 & 1024 & 2048 & 7168 & 20480 \\ \hline 
 64
&262144 & 117649 & 4096 & 8192 & 20480 & 61440 \\ \hline 
 128
&2097152 & 823543 & 16384 & 16384 & 65536 & 180224 \\ \hline 
 256
&16777216 & 5764801 & 65536 & 65536 & 196608 & 589824 \\ \hline 
 512&134217728 & 40353607 & 262144 & 262144 & 524288 & 1835008 \\ \hline
\end{tabular}
\caption{Number of multiplications required for plaintext matrix multiplication using schoolbook approach, Strassen's algorithm and Cussen's algorithm for random square matrices of dimension $n \in \{2^3, 2^4, \cdots, 2^9\}$ with element bit-widths $t \in \{4, 8, 12, 16\}$ (data for Figure \ref{fig:cussen_plaintext_matrix_python}).}
\label{table:appendix:cussen_plaintext_matrix_python_mul}
\end{table}

\begin{table}[!ht]
\centering
\renewcommand{\arraystretch}{1.25}
\scriptsize
\begin{tabular}{|c|c|c|c|c|c|c|} \hline       
 $n$
&Schoolbook & Strassen & Cussen $(t = 4)$ & Cussen $(t = 8)$ & Cussen $(t = 12)$ & Cussen $(t = 16)$ \\ \hline 
 8
&448 & 1674 & 1152 & 1792 & 1984 & 1984 \\ \hline 
 16
&3840 & 12870 & 7424 & 12800 & 15616 & 16128 \\ \hline 
 32
&31744 & 94698 & 48128 & 81920 & 112640 & 128000 \\ \hline 
 64
&258048 & 681318 & 327680 & 557056 & 782336 & 962560 \\ \hline 
 128
&2080768 & 4842954 & 2359296 & 3899392 & 5308416 & 6799360 \\ \hline 
 256
&16711680 & 34195590 & 17825792 & 27787264 & 36831232 & 48627712 \\ \hline 
 512&133955584 & 240548778 & 138412032 & 192937984 & 271056896 & 338690048 \\ \hline
\end{tabular}
\caption{Number of additions required for plaintext matrix multiplication using schoolbook approach, Strassen's algorithm and Cussen's algorithm for random square matrices of dimension $n \in \{2^3, 2^4, \cdots, 2^9\}$ with element bit-widths $t \in \{4, 8, 12, 16\}$ (data for Figure \ref{fig:cussen_plaintext_matrix_python}).}
\label{table:appendix:cussen_plaintext_matrix_python_add}
\end{table}

\begin{table}[!ht]
\centering
\renewcommand{\arraystretch}{1.25}
\begin{tabular}{|c|c|c|c|} \hline
  $t$&$n$
&Schoolbook & Proposed \\ \hline
\multirow{7}{*}{4}&8
&128 & 38 \\ 
 &16
&256 & 44 \\ 
 &32
&512 & 48 \\ 
 &64
&1024 & 50 \\ 
 &128
&2048 & 50 \\ 
 &256
&4096 & 50 \\ 
 &512&8192 & 50 \\ \hline 
\multirow{7}{*}{8}&8
&256 & 170 \\ 
  &16
&512 & 166 \\ 
  &32
&1024 & 162 \\ 
  &64
&2048 & 210 \\ 
  &128
&4096 & 254 \\ 
  &256
&8192 & 370 \\ 
  &512&16384 & 482 \\ \hline 
\multirow{7}{*}{12}&8
&384 & 432 \\ 
  &16
&768 & 524 \\ 
  &32
&1536 & 494 \\ 
  &64
&3072 & 496 \\ 
  &128
&6144 & 586 \\ 
  &256
&12288 & 758 \\ 
  &512&24576 & 1142 \\ \hline 
\multirow{7}{*}{16}&8
&512 & 560 \\ 
  &16
&1024 & 1056 \\ 
  &32
&2048 & 1468 \\ 
  &64
&4096 & 1304 \\ 
  &128
&8192 & 1280 \\ 
  &256
&16384 & 1550 \\ 
  &512&32768 & 2010 \\ \hline
\end{tabular}
\caption{Equivalent number of elliptic curve point additions required for plaintext vector and ciphertext scalar multiplication using schoolbook approach and proposed approach based on Cussen's algorithm for random vectors of length $n \in \{2^3, 2^4, \cdots, 2^9\}$ with element bit-widths $t \in \{4, 8, 12, 16\}$ (data for Figure \ref{fig:cussen_ciphertext_vector_python}).}
\label{table:appendix:cussen_ciphertext_vector_python}
\end{table}

\begin{table}[!ht]
\centering
\renewcommand{\arraystretch}{1.25}
\begin{tabular}{|c|c|c|c|c|} \hline
  $t$&$n$
&Schoolbook & Strassen & Proposed \\ \hline
 \multirow{7}{*}{4} &8
&9088 & 7906 & 3328 \\ 
  &16
&73216 & 57006 & 18944 \\ 
  &32
&587776 & 405698 & 112640 \\ 
  &64
&4710400 & 2866510 & 720896 \\ 
  &128
&37715968 & 20172066 & 4980736 \\ 
  &256
&301858816 & 141630446 & 36700160 \\ 
  &512&2415394816 & 993117058 & 281018368 \\ \hline
\multirow{7}{*}{8}  &8
&17280 & 13394 & 11776 \\ 
  &16
&138752 & 95422 & 50176 \\ 
  &32
&1112064 & 674610 & 229376 \\ 
  &64
&8904704 & 4748894 & 1376256 \\ 
  &128
&71270400 & 33348754 & 8323072 \\ 
  &256
&570294272 & 233867262 & 57671680 \\ 
  &512&4562878464 & 1638774770 & 394264576 \\ \hline
 \multirow{7}{*}{12} &8
&25472 & 18882 & 28544 \\ 
  &16
&204288 & 133838 & 141824 \\ 
  &32
&1636352 & 943522 & 569344 \\ 
  &64
&13099008 & 6631278 & 2547712 \\ 
  &128
&104824832 & 46525442 & 13762560 \\ 
  &256
&838729728 & 326104078 & 83099648 \\ 
  &512&6710362112 & 2284432482 & 567279616 \\ \hline
\multirow{7}{*}{16}  &8
&33664 & 24370 & 36736 \\ 
  &16
&269824 & 172254 & 278016 \\ 
  &32
&2160640 & 1212434 & 1566720 \\ 
  &64
&17293312 & 8513662 & 5857280 \\ 
  &128
&138379264 & 59702130 & 25133056 \\ 
  &256
&1107165184 & 418340894 & 135004160 \\ 
  &512&8857845760 & 2930090194 & 794820608 \\ \hline
\end{tabular}
\caption{Equivalent number of elliptic curve point additions required for plaintext-ciphertext matrix multiplication \texttt{PC-MM} using schoolbook approach, Strassen's algorithm and proposed approach based on Cussen's algorithm for random square matrices of dimension $n \in \{2^3, 2^4, \cdots, 2^9\}$ with element bit-width $t \in \{4, 8, 12, 16\}$ (data for Figure \ref{fig:cussen_ciphertext_matrix_python}).}
\label{table:appendix:cussen_ciphertext_matrix_python}
\end{table}

\clearpage

\subsection{Measurements on Raspberry Pi 5}

\begin{table}[!ht]
\centering
\begin{tabular}{|c|c|c|c|} \hline
  $t$&$n$&Schoolbook & Proposed \\ \hline
 \multirow{7}{*}{4} &8&929.364 & 250.138 \\ 
  &16&1853.07 & 261.694 \\ 
  &32&3700.836 & 296.418 \\ 
  &64&7400.958 & 318.018 \\ 
  &128&14837.766 & 335.144 \\ 
  &256&29592.7 & 353.944 \\ 
  &512&59153.174 & 399.094 \\ \hline
 \multirow{7}{*}{8} &8&1194.568 & 723.292 \\ 
  &16&2386.114 & 588.268 \\ 
  &32&4765.696 & 551.734 \\ 
  &64&9526.752 & 615.192 \\ 
  &128&19047.554 & 765.108 \\ 
  &256&38086.59 & 1071.834 \\ 
  &512&76327.824 & 1473.63 \\ \hline
 \multirow{7}{*}{12} &8
&1460.39 & 1472.122 \\ 
  &16
&2917.274 & 1905.484 \\ 
  &32
&5830.14 & 1563.06 \\ 
  &64
&11652.792 & 1458.826 \\ 
  &128
&23298.534 & 1583.76 \\ 
  &256
&46590.522 & 1978.638 \\ 
  &512&93181.214 & 3044.194 \\ \hline
 \multirow{7}{*}{16} &8
&1729.608 & 1814.146 \\ 
  &16
&3445.01 & 3502.018 \\ 
  &32
&6884.118 & 4633.108 \\ 
  &64
&13775.728 & 3949.804 \\ 
  &128
&27523.884 & 3692.986 \\ 
  &256
&55047.72 & 4172.836 \\ 
  &512&110086.372 & 5272.712 \\ \hline
\end{tabular}
\caption{Time taken (in micro seconds) for plaintext vector and ciphertext scalar multiplication using schoolbook approach and proposed approach based on Cussen's algorithm for random vectors of length $n \in \{2^3, 2^4, \cdots, 2^9\}$ with element bit-width $t \in \{4, 8, 12, 16\}$ (data for Figure \ref{fig:cussen_ciphertext_vector_rpi}).}
\label{table:appendix:cussen_ciphertext_vector_rpi}
\end{table}

\begin{table}[!ht]
\centering
\begin{tabular}{|c|c|c|c|c|} \hline  
 $n$& $t = 4$ & $t = 8$ & $t = 12$ & $t = 16$ \\ \hline  
8 & 3.715 & 1.652 & 0.992 & 0.953 \\ \hline  
16 & 7.081 & 4.056 & 1.531 & 0.984 \\ \hline  
32 & 12.485 & 8.638 & 3.73 & 1.486 \\ \hline  
64 & 23.272 & 15.486 & 7.988 & 3.488 \\ \hline  
128 & 44.273 & 24.895 & 14.711 & 7.453 \\ \hline  
256 & 83.608 & 35.534 & 23.547 & 13.192 \\ \hline  
512 & 148.219 & 51.796 & 30.609 & 20.879 \\ \hline 
\end{tabular}
\caption{Speedup observed for plaintext vector and ciphertext scalar multiplication using proposed approach based on Cussen's algorithm compared to schoolbook approach for random vectors of length $n \in \{2^3, 2^4, \cdots, 2^9\}$ with element bit-widths $t \in \{4, 8, 12, 16\}$ (data for Figure \ref{fig:speedup_vector}).}
\label{table:appendix:speedup_vector_rpi}
\end{table}

\begin{table}[!ht]
\centering
\begin{tabular}{|c|c|c|c|c|} \hline
 $t$ &$n$
&Schoolbook & Strassen & Proposed \\ \hline
 \multirow{7}{*}{4} &8
&0.06 & 0.05 & 0.02 \\ 
  &16
&0.49 & 0.4 & 0.09 \\ 
  &32
&3.93 & 2.95 & 0.43 \\ 
  &64
&31.43 & 21.73 & 2.46 \\ 
  &128
&251.74 & 158.82 & 14.31 \\ 
  &256
&2017.64 & 1161.86 & 95.5 \\ 
  &512&16139.04 & 8480.31 & 687.85 \\ \hline
 \multirow{7}{*}{8} &8
&0.08 & 0.07 & 0.05 \\ 
  &16
&0.63 & 0.48 & 0.18 \\ 
  &32
&5.02 & 3.51 & 0.7 \\ 
  &64
&40.13 & 25.66 & 3.65 \\ 
  &128
&321.82 & 186.59 & 20.96 \\ 
  &256
&2571.89 & 1355.63 & 139.56 \\ 
  &512&20577.63 & 9824.45 & 932.97 \\ \hline
 \multirow{7}{*}{12} &8 
&0.1 & 0.08 & 0.1 \\ 
  &16 
&0.76 & 0.56 & 0.5 \\ 
  &32 
&6.11 & 4.07 & 1.71 \\ 
  &64 
&48.92 & 29.52 & 7.13 \\ 
  &128 
&391.16 & 213.63 & 34.53 \\ 
  &256 
&3134.81 & 1544.58 & 198.54 \\ 
  &512 &25079.63 & 11292.73 & 1305.7 \\ \hline
 \multirow{7}{*}{16} &8 
&0.11 & 0.09 & 0.12 \\ 
  &16 
&0.9 & 0.64 & 0.91 \\ 
  &32 
&7.19 & 4.63 & 4.93 \\ 
  &64 
&57.61 & 33.43 & 17.14 \\ 
  &128 
&461.59 & 241.41 & 69.5 \\ 
  &256 
&3687.85 & 1759.8 & 334.51 \\ 
  &512 &29503.38 & 12489.93 & 1862.09 \\ \hline
\end{tabular}
\caption{Time taken (in seconds) for plaintext-ciphertext matrix multiplication \texttt{PC-MM} using schoolbook approach, Strassen's algorithm and proposed approach based on Cussen's algorithm for random square matrices of dimension $n \in \{2^3, 2^4, \cdots, 2^9\}$ with element bit-width $t \in \{4, 8, 12, 16\}$ (data for Figure \ref{fig:cussen_ciphertext_matrix_rpi}).}
\label{table:appendix:cussen_ciphertext_matrix_rpi}
\end{table}

\begin{table}[!ht]
\centering
\begin{tabular}{|c|c|c|c|} \hline
 $t$ &$n$& Schoolbook vs Proposed & Strassen vs Proposed \\ \hline
 \multirow{7}{*}{4} &8 & 3.68 & 3.23 \\ 
 &16 & 5.54 & 4.52 \\ 
 &32 & 9.05 & 6.8 \\ 
 &64 & 12.79 & 8.84 \\ 
 &128 & 17.59 & 11.1 \\ 
 &256 & 21.13 & 12.17 \\ 
 &512 & 23.46 & 12.33 \\ \hline
 \multirow{7}{*}{8} &8 & 1.6 & 1.33 \\ 
 &16 & 3.5 & 2.67 \\ 
 &32 & 7.2 & 5.04 \\ 
 &64 & 11 & 7.03 \\ 
 &128 & 15.36 & 8.9 \\ 
 &256 & 18.43 & 9.71 \\ 
 &512 & 22.06 & 10.53 \\ \hline
 \multirow{7}{*}{12} &8 & 0.99 & 0.79 \\ 
 &16 & 1.53 & 1.12 \\ 
 &32 & 3.58 & 2.38 \\ 
 &64 & 6.86 & 4.14 \\ 
 &128 & 11.33 & 6.19 \\ 
 &256 & 15.79 & 7.78 \\ 
 &512 & 19.21 & 8.65 \\ \hline
 \multirow{7}{*}{16} &8 & 0.95 & 0.74 \\ 
 &16 & 0.99 & 0.7 \\ 
 &32 & 1.46 & 0.94 \\ 
 &64 & 3.36 & 1.95 \\ 
 &128 & 6.64 & 3.47 \\ 
 &256 & 11.02 & 5.26 \\ 
 &512 & 15.84 & 6.71 \\ \hline
\end{tabular}
\caption{Speedups observed for plaintext-ciphertext matrix multiplication \texttt{PC-MM} using proposed approach based on Cussen's algorithm compared to schoolbook approach and Strassen's algorithm for random square matrices of dimension $n \in \{2^3, 2^4, \cdots, 2^9\}$ with element bit-width $t \in \{4, 8, 12, 16\}$ (data for Figure \ref{fig:speedup_matrix}).}
\label{table:appendix:speedup_matrix_rpi}
\end{table}

\end{document}